\documentclass[lettersize,journal]{IEEEtran}
\usepackage[caption=false,font=normalsize,labelfont=sf,textfont=sf]{subfig}
\IEEEoverridecommandlockouts
\usepackage[compact]{titlesec} 
\usepackage{amsmath,amsfonts}
\usepackage{booktabs}
\usepackage{tabularx}
\usepackage{multirow}
\usepackage{color}
\usepackage{float}
\usepackage{cite}
\usepackage{enumitem}  
\usepackage[utf8]{inputenc}
\usepackage{lmodern}
\usepackage{tabularx}
\usepackage{csquotes}
\usepackage[justification=centering]{caption}
\usepackage{blindtext}
\usepackage{wrapfig}
\usepackage{graphicx}
\usepackage[bottom]{footmisc}
\DeclareGraphicsExtensions{.eps,.pdf,.gif,.png,.jpg}
\usepackage{amssymb}
\usepackage{amsthm}
\usepackage{array}
\usepackage{textcomp}
\usepackage{stfloats}
\usepackage{algorithm,algorithmic}
\usepackage{booktabs}
\usepackage{graphicx}
\usepackage{epstopdf}\usepackage[utf8]{inputenc}
\usepackage[hyphens,spaces,obeyspaces]{url}
\usepackage[english]{babel}
\usepackage{verbatim} 
\usepackage{lineno,nohyperref}
\usepackage{epstopdf} %converting to PDF
\modulolinenumbers[5]

\RequirePackage{filecontents}
\usepackage{ragged2e}
\usepackage{url} 
\usepackage[T1]{fontenc}
\usepackage[utf8]{inputenc}
\usepackage{microtype} %remove empy space in URL

\newcommand{\vect}[1]{\boldsymbol{#1}}
\newcommand*{\email}[1]{#1}
\PassOptionsToPackage{hyphens}{url}
\newtheoremstyle{mystyle}%                % Name
{}%                                     % Space above
{}%                                     % Space below
{\itshape}%                                     % Body font
{}%                                     % Indent amount
{\bfseries}%                            % Theorem head font
{.}%                                    % Punctuation after theorem head
{ }%                                    % Space after theorem head, ' ', or \newline
{\thmname{#1}\thmnumber{ #2}\thmnote{ (#3)}}%                                     % Theorem head spec (can be left empty, meaning `normal')

\theoremstyle{mystyle}
\newtheorem{remark}{Remark}

\newcounter{subassumption}[asu]

\makeatletter
\renewcommand{\p@subassumption}{\theasu}% Counter prefix.
\makeatother
\usepackage{amsthm}
\usepackage{xpatch,amsthm}
\makeatletter
\xpatchcmd{\@thm}{\fontseries\mddefault\upshape}{}{}{} % same font as thm-header
\makeatother

\makeatother
\usepackage{cite}
\usepackage{amsmath,amssymb,amsfonts}
\usepackage{algorithmic}
\usepackage{graphicx}
\usepackage{textcomp}
\usepackage{xcolor}
\def\BibTeX{{\rm B\kern-.05em{\sc i\kern-.025em b}\kern-.08em
		T\kern-.1667em\lower.7ex\hbox{E}\kern-.125emX}}
\DeclareMathOperator{\E}{\mathbb{E}}
\DeclareMathOperator{\proj}{proj}
\begin{document}

\title{Age of Processing-Based Data Offloading for Autonomous Vehicles in Multi-RATs Open RAN\\
	\thanks{}}
\author{Anselme~Ndikumana,~\IEEEmembership{Member,~IEEE,~}
	Kim~Khoa~Nguyen,~\IEEEmembership{Senior~Member,~IEEE,~}\\
	and~Mohamed~Cheriet,~\IEEEmembership{Senior~Member,~IEEE,~}
	\IEEEcompsocitemizethanks{
		\IEEEcompsocthanksitem Anselme Ndikumana, Kim Khoa Nguyen, and Mohamed Cheriet are  with Synchromedia Lab, École de
		Technologie Supérieure, Université du Québec, QC, Canada, E-mail: (\email{anselme.ndikumana.1@ens.etsmtl.ca; kim-khoa.nguyen@etsmtl.ca; Mohamed.Cheriet@etsmtl.ca}).
		%\IEEEcompsocthanksitem 	``The authors thank Mitacs, Ciena, and ENCQOR for funding this research under the IT13947 grant''.
	}}
\maketitle

\begin{abstract}
	Today,  vehicles use smart sensors to collect data from the road environment. This data is often processed onboard of the vehicles, using expensive hardware.  Such onboard processing increases the vehicle's cost, quickly drains its battery, and exhausts its computing resources. Therefore, offloading tasks onto the cloud is required. Still, data offloading is challenging due to low latency requirements for safe and reliable vehicle driving decisions.  Moreover, age of processing was not considered in prior research dealing with low-latency offloading for autonomous vehicles. This paper proposes an age of processing-based offloading approach for autonomous vehicles using unsupervised machine learning, Multi-Radio Access Technologies (multi-RATs), and Edge Computing in  Open Radio Access Network (O-RAN). We design a collaboration space of edge clouds to process data in proximity to autonomous vehicles. To reduce the variation in offloading delay, we propose a new communication planning approach that enables the vehicle to optimally preselect the available RATs such as Wi-Fi, LTE, or 5G  to offload tasks to edge clouds when its local resources are insufficient. We formulate an optimization problem for age-based offloading that minimizes elapsed time from generating tasks and receiving computation output.  To handle this non-convex problem, we develop a surrogate problem. Then, we use the Lagrangian method to transform the surrogate problem to unconstrained optimization problem and apply the dual decomposition method. The simulation results show that our approach significantly minimizes the age of processing in data offloading with 90.34\%  improvement over similar method.
\end{abstract}


\section{Introduction}
\label{sec:introduction}
\IEEEPARstart{C}{ellular} vehicle-to-everything (C-V2X) has recently been introduced in 5G to enable low-latency and high-reliability vehicular communications, ultimately supporting autonomous driving. C-V2X integrates V2V (Vehicle-to-Vehicle), V2I (Vehicle-to-Infrastructure),  V2P (Vehicle-to-Pedestrian), and V2N (Vehicle-to-Network) by leveraging cellular network infrastructure \cite{chen2020vision}. At the same time, in cellular network, to support lower latency communication in Radio Access Network (RAT), the Open Radio Access Network (O-RAN) architecture  has recently been proposed   \cite{allianceORANUseCases}. The O-RAN architecture enables the intelligence and openness of RAN, and it can achieve nearly real-time optimization of RAN resources using Machine Learning (ML) algorithms implemented in the Near Real-Time RAN Intelligent Controller (Near-RT RIC). O-RAN architecture enables collecting and accessing historical traffic and handover data in Near-RT RIC. Near-RT RIC can use ML to detect the network and handover anomalies and ensure continuous and reliable connectivity for autonomous driving.  The Near- RT RIC is interfaced with O-RAN Central Unit Control Plane (O-CU-CP) and O-RAN Central Unit User Plane (O-CU-UP) at edge cloud called ``Open Cloud (O-Cloud)''. Also, in O-RAN, Non-Real-Time RAN Intelligent Controller (Non-RT RIC) enables ML functionalities for policy-based guidance of applications and features. Therefore, we consider O-RAN and C-V2X as key enabling communication technologies toward low-latency communications for autonomous driving. Furthermore, to enable lower latency in the presence of multiple Radio Access Technologies (multi-RATs) such as Wi-Fi, LTE, or 5G, the 3rd Generation Partnership Project (3GPP) proposed a Non-3GPP Interworking Function (N3IWF). N3IWF allows controlling various RATs in a unified manner in 5G core. Also, the 3GPP  defines Access Traffic Steering, Switching, and Splitting (ATSSS) functionality that allows traffic steering, switching, and splitting for multi-RATs environments \cite{ETSITS}. Therefore, to improve reliability and lower latency of autonomous driving in multi-RATs,  we consider redundant user planes with  ATSSS, N3IWF, and User Plane Functions (UPFs) at the edge clouds. Such consideration allows data of autonomous vehicles in 5G, LTE, and Wi-Fi to be routed to User Plane Function (UPF) directly and via N3IWF.

In addition to vehicular communication networks, the automotive and transportation industries have recently made enormous investments in autonomous driving and Artificial Intelligence (AI) \cite{ma2020artificial} to develop Intelligent Transportation System (ITS) \cite{ndikumana2020deep}. In 2019, the global autonomous vehicle market was valued at USD 24.1 billion. For the forecast period, 2020-2025,   the autonomous vehicle market expected a Compound Annual Growth Rate (CAGR) of 18.06\% \cite{GlobeNewswire1}. According to \cite{yaqoob2019autonomous}, autonomous driving using AI in ITS can help in providing reliable transportation services by eliminating many accidents that could be caused by human errors. The ITS provides information and recommends driving decisions to drivers and self-driving cars. This requires collaborative sensing and information exchange between vehicles and infrastructure, where communication network is a substantial enabling element of ITS \cite{bagheri20215g}. Therefore, ITS requires a combination of various cutting-edge information using AI, computation, and communication technologies for traffic signal control, route optimization, emergency driving assistance, intelligent parking, etc \cite{zhou2021intelligent}. Consequently, ITS should use massive amounts of data from various sources such as vehicles, pedestrians, passengers, and roadside units\cite{zhu2018big}. According to \cite{Accenture}, an autonomous vehicle alone (level $2$ and above of
driving automation) can generate between $4$ and $10$ terabytes of data per day, depending on the number of mounted sensors. % Handling such massive data using vehicle On-Board Units (OBU) can drain the battery quickly.%clouds need to be evolved to support vehicles in ITS. 
%Therefore, the critical challenge in ITS is to have enough energy, computing, and communication resources to handle massive generated data from various sources. This generated data requires real-time analytics to guarantee low latency communication and safe and reliable transportation.
In practice, the most critical computation challenges faced by autonomous vehicles when processing collected data from sensors are:
\begin{itemize}
	\item
	In the autonomous vehicle, On-Board Units (OBU) does all processing/computation of collected data from sensors.  Thus, OBU consumes a lot of battery energy leading to a shorter battery and OBU lifetime \cite{tang2020container}. Also, handling all compute-intensive tasks in the autonomous vehicle may exceed the available resource capacity.
	\item  The time for getting computational results is critical for autonomous driving. In the worst-case, computation time may exceed the deadline bound to make safe and reliable autonomous driving decisions.
	\item To support OBU, autonomous vehicles may offload tasks to edge clouds using multi-RATs. Therefore, data from multiple autonomous vehicles may rapidly reach edge cloud with mixed finite and infinite flows with varying rates. Considering that each edge cloud works independently,  required computation resources may exceed the available resources of one edge cloud. 
	\item 
	Offloading data to the edge cloud depends on the network status. However, network status frequently changes over time and causes fluctuation of end-to-end latency \cite{cui2020offloading, ndikumana2023federated}. 
	\item 
	Offloading data in the presence of multi-RATs involves multiple handovers due to the vehicle's connection in motion and high mobility. Multiple handovers affect computation and communication delay.
\end{itemize}

In this work, we consider computation and communication delay in handling data from vehicles. We opt Age of Processing (AoP), where AoP comes from Age of Information (AoI). AoI has been proposed to measure the status freshness of the environment \cite{yates2021age}. AoI captures the time elapsed from status being generated at the source node to the latest status update at the destination node. However,  we can get the status information after performing some data processing, i.e., computation of collected data. To include computation time in the AoI,  we consider AoP for vehicle data offloading to edge cloud in multi-RATs environment. The AoP considers the time elapsed from generating task to the time of receiving computation output. AoP has been applied in data sampling, offloading, and processing  for real-time Internet of Things (IoT) applications \cite{li2021age}. To the best of our knowledge, this work is the first that considers AoP in autonomous driving. We propose an AoP-based data offloading for autonomous vehicles in multi-RATs Open RAN to tackle the aforementioned challenges. Our main contributions are summarized as follows:
\begin{itemize}
	\item
	We propose a Collaboration Space (CS) of edge clouds to compute tasks as close as possible to autonomous vehicles for minimizing AoP. The CS is defined using Affinity Propagation (AP)- an unsupervised ML algorithm implemented in O-RAN controllers. AP allows putting edge clouds in CSs based on their similarity, responsibility, and availability.
	\item
	We propose a communication planning approach to reduce unpredictable variation in offloading delay.  The vehicle can preselect appropriate RATs available in its route before its road trip. Then, the vehicle can choose a suitable RAT among the preselected RATs to immediately start offloading tasks when its local computing resource is insufficient.
	\item We formulate an optimization approach that jointly optimizes communication and computation models to minimize the AoP of autonomous vehicles. The formulated problem is shown to be non-convex and computationally intractable. To handle it, we develop a surrogate and upbound problem of the original problem. Then, we transform the surrogate problem to an unconstrained optimization problem using the Lagrangian method and apply dual decomposition  to solve it.
\end{itemize}
%\textcolor{blue}{The authors in \cite{tang2020container} proposed offloading autonomous driving services for edge computing to minimize response delay. In \cite{cui2020offloading}, authors discussed a new approach to offload computation-intensive autonomous driving tasks to roadside units and cloud for executions. The proposed system aims to minimize response time. In \cite{liu2022mobility}, mobility-aware task offloading was formulated as an optimization problem that minimizes execution time. In terms of AoI, the authors in \cite{sorkhoh2021optimizing} optimize information freshness of autonomous vehicles in cooperative Multi-access Edge Computing (MEC). The priors works do not consider the offloading approach where the vehicles navigate the areas covered by multi-RATs. Also, the AoI-based offloading approach does not capture computation time in AoI. The offloading problem that considers AoP in  multi-RATs and multi-edge computing environments was not considered in the literature.} 

The rest of this paper is structured as follows: we discuss related work in Section  \ref{sec:Related Works}.  Section  \ref{sec:system-model} presents our system model, and Section \ref{sec:FederatedLearningl} discusses age-based task offloading. In Section \ref{sec:Problem_Formulation}, we present our problem formulation and proposed solution.  Section \ref{sec:PerformanceEvaluation} presents our performance evaluation, and we conclude the paper in Section \ref{sec:Conclusion}.

\section{Related Work} 
\label{sec:Related Works}
We classify the existing related work into two  categories:
($i$) offloading and autonomous vehicles, and  ($ii$)  offloading and age of information.

\emph{($i$) Offloading and autonomous vehicles:}  The authors in \cite{silva2021computing} highlighted the need for a highly efficient, fast, and integrated network supporting data offloading. Specifically, Multi-RATS can increase network capacity and throughput. The related works in \cite{ndikumana2017collaborative,ndikumana2019joint} discussed communication and computation approaches for task offloading in multi-access edge computing. Rather than considering binary offloading available in \cite{zhu2021multiobjective, ndikumana2017collaborative,ndikumana2019joint}, where each task is either computed locally or entirely offloaded to the edge cloud, the authors in \cite{bi2020energy} proposed energy efficiency partial computation offloading  in which a task is divisible for being executed parallelly in different locations.  
Inspired by artificial intelligence, the authors in \cite{ning2020intelligent, ndikumana2020deep} proposed offloading and caching approaches that enable vehicle to Road Side Unit (RSU) offloading.
%The authors in \cite{tang2020container} proposed container-based offloading for autonomous driving and formulated the offloading problem as a multidimensional knapsack problem that aims to maximize offloading utility. 
Using edge computing,  the authors in \cite{cui2020offloading} proposed offloading autonomous driving services.  However, in their proposed approach,  the authors consider only a single edge server. Since many autonomous vehicles may offload tasks to the edge server at the same time, their demands may surpass the capacity of a single edge server. %Also, offloading to remote clouds is not appropriate for autonomous driving due to stringent response time  requirements.
The authors in \cite{prathiba2021federated} proposed a radio resource management approach for offloading tasks to edge clouds using 6G V2X communications. They used Federated Q-Learning to allocate and utilize the available radio resources efficiently. 
In \cite{li2021federated}, the authors presented a data sharing approach in the vehicular edge network that minimizes transmission latency in data sharing. To handle the formulated problem, they used Q-network and federated learning approaches to
ensure efficient and secure data sharing among multi-access edge computing (MEC) servers.
In \cite{guo2020intelligent}, the authors defined an optimization problem for vehicle offloading and communications selection decisions in the MEC environment. They used a deep Q-learning approach to find the optimal solution. The authors in \cite{khayyat2020advanced}  minimize delay and energy consumption for multilevel offloading in a vehicular network. They used distributed Deep Q-learning algorithm to handle the formulated problem by maximizing reward. However, the aforementioned ML-based approaches do not provide a computational complexity analysis to prove their applicability in the driving environment. Also, the proposed approaches cannot easily be implemented in O-RAN due to distributed O-RAN elements.

\emph{($ii$) Offloading and age of information:} 
The authors in \cite{modina2020joint} proposed joint traffic offloading and AoI control for data collected by IoT devices in the smart city. They presented a price-based mechanism using AoI to minimize the costs of service providers. Similarly,  using AoI, the authors in \cite{song2019age} proposed the age of task for evaluating task computation of mobile edge computing systems in terms of temporal values. The authors considered a system with a single mobile edge computing server and a single mobile device. In  \cite{modina2020joint, song2019age}, the authors do not consider fronthaul and midhaul links of 5G networks in their AoI models. Also, in intelligent systems such as video surveillance, status information can only be available after some computation, which takes time. Thus, the time required for data processing affects the status freshness; the authors in  \cite{li2021age} proposed age-driven status sampling and processing offloading that aims to minimize AoP for IoT.

In general, prior work has not yet addressed the problem of offloading for autonomous vehicles in an environment consisting of both Multi-RATs and multi-edge computing. Also, O-RAN is a newly introduced architecture, therefore prior work has not tackled the task offloading problem in Multi-RATs using 5G O-RAN architecture. No prior work has presented a joint offloading and communication planning problem that uses AoP and leverages O-RAN controllers. To this end, our proposed approach has several novelties over these prior approaches, including: ($i$)  modeling  AoP of offloaded tasks from autonomous vehicles; ($ii$) defining CS of edge clouds to minimize AoP; ($iii$) proposing a new communication planning approach that enables the autonomous vehicle to preselect RATs available in its route for task offloading; ($iv$) designing an offloading approach in O-RAN environment that considers redundant user planes of multiple  User Plane Function (UPF) and  RATs at the edge to improve reliability and lower latency of autonomous driving.

%\textcolor{blue}{To this end, our proposed approach has several novelties over these prior approaches, including:  ($i$) joint offloading and communication planning problem that leverages O-RAN controllers in distributing tasks to multiple edge clouds; ($ii$)  investigating  AoP of offloaded tasks from autonomous vehicles; ($iii$) the CS of edge clouds that minimizes AoP; ($iv$) a new communication planning approach that enables the autonomous vehicle to preselect RATs available in its route for task offloading; ($v$) an offloading approach in O-RAN environment considers redundant user planes of multiple  User Plane Function (UPF) and  RATs at the edge to improve reliability and lower latency of autonomous driving.}
\begin{figure}[t]
	\centering
	\includegraphics[width=1.0\columnwidth]{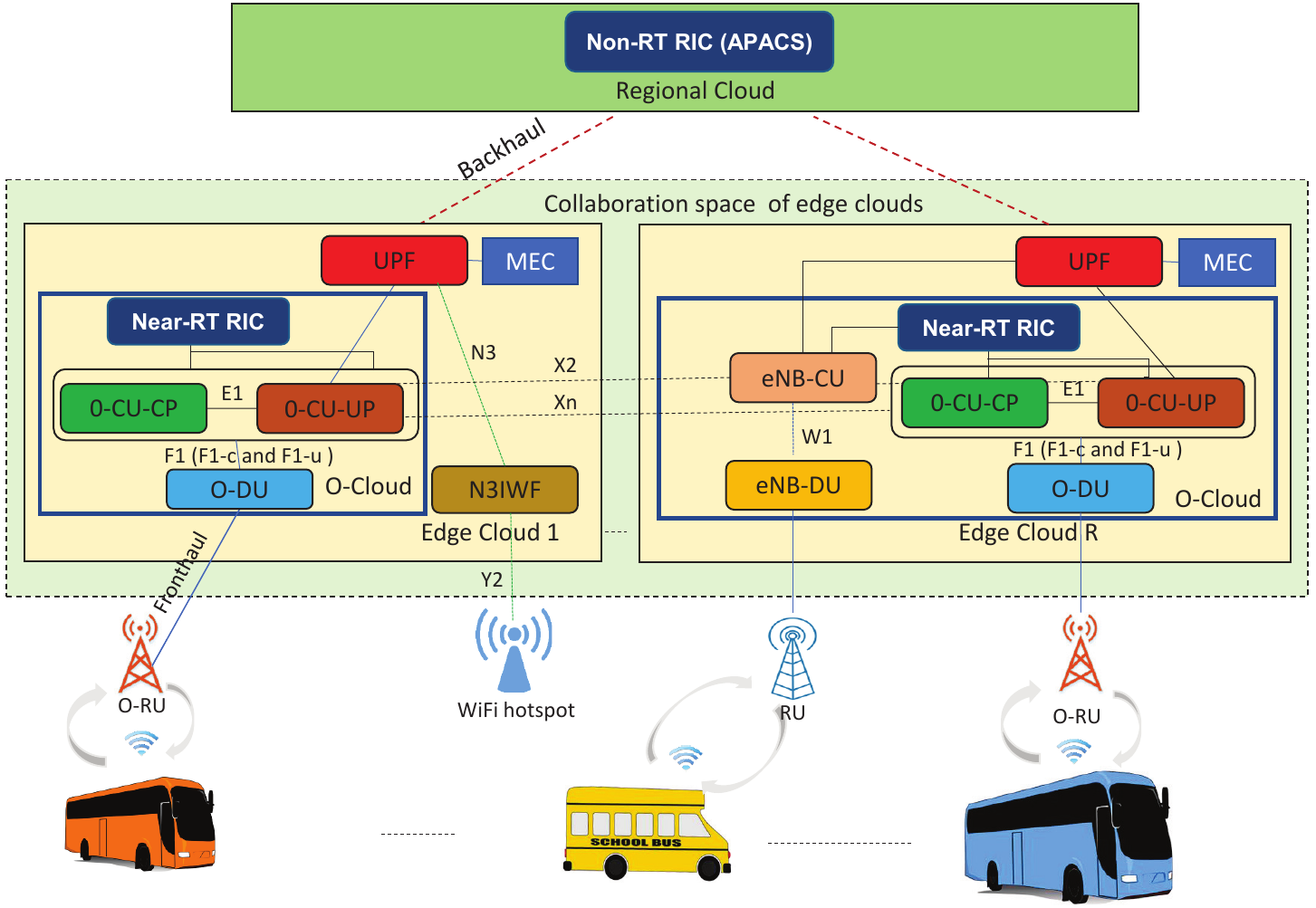}
	\caption{Illustration of our system model.}
	\label{fig:SystemModel}
\end{figure}
\begin{table}[t]
	\caption{Summary of key notations.}
	\label{tab:table1}
	\begin{tabular}{ll}
		\toprule
		Notation & Definition\\
		\midrule
		$\mathcal{R}$ & Set of edge clouds, $|\mathcal{R}|= R$\\
		$\mathcal{V}$ & Set of autonomous vehicles,  $|\mathcal{V}|= V$\\
		$\mathcal{S}$ & Set of RATs,  $|\mathcal{S}|= S$\\
		$\mathcal{O}$ & Set of routers/switches\\
		$\mathcal{W}$ & Set of collaboration spaces,  $|\mathcal{W}|= W$\\
		$\mathcal{E}$ & Set of links, $|\mathcal{E}|= E$\\ 
		$\vect{\Phi}$ & Responsibility matrix \\
		$\vect{\Theta}$& Availability matrix \\
		$Z$ & Function that computes the similarity\\		
		$K^v_i$ & Time of new status update $i$ sampling at $v \in \mathcal{V}$\\
		$A^{av}$ & Average AoP  for autonomous vehicle $v$ \\
		$u_v(t)$ & Freshest status update at autonomous vehicle $v$\\ 
		$x_v^{s\rightarrow r}$ & Offloading variable for autonomous vehicle $v$\\ 
		%		$\alpha_{v}$ & autonomous vehicle  status parameter\\ 
		%		$\varphi_v$	& Average waiting time of task $T_{v}$ \\
		&	for being executed at autonomous vehicle $v$\\ 
		$\omega_{r, RC}$ & Capacity of wired link between EC $r$ and RC\\
		$\omega_{v, s}$ & Capacity of wireless link between UE $v$ and RAT $s$\\
		$T_{v}$ & Task of autonomous vehicle $v \in \mathcal{V}$\\
		$\tau^\textrm{loc}_{v}$ & Local computational delay of autonomous vehicle $v$\\
		$\tau^\textrm{off}_{v}$ & Total offloading delay of autonomous vehicle $v \in \mathcal{V}$\\ 
		$p_r$ & Computation capability of  EC	$r \in \mathcal{R}$\\
		$p_v$ & Computation capability  of each  vehicle $v \in \mathcal{V}$\\
		$L^v_i$& Total offloading and computation time \\ 
		&	of status update $i$ for autonomous vehicle $v$\\  
		\bottomrule
	\end{tabular}
\end{table}
%---------------------------------------------------------------

\section{System Model}
\label{sec:system-model}
Fig. \ref{fig:SystemModel} illustrates the system model of our proposed AoP-based offloading approach for autonomous vehicles. %This figure does not cover the 5G core, the internal structure of the regional cloud, and some O-RAN interfaces for easy visualization. Furthermore, 
The summary of our key notations in the proposed model is available in Table \ref{tab:table1}.

\emph{Autonomous vehicles:}  We consider a set  $\mathcal{V}$ of autonomous vehicles. Each vehicle has an application that generates compute-intensive tasks, where $T_{v} = (s{d_v},\tilde{\tau}_{v}, \tilde{z}_{v})$  is a computation task of  vehicle $v$. In  $T_{v}$, $s{d_v}$ is the size of input data $d_v$ in bits. We denote by $\tilde{\tau}_{v}$  the task computation deadline, and by $\tilde{z}_{v}$ the computation workload.  Each autonomous vehicle $v \in \mathcal{V}$ has an OBU with computational capacity $P_v$, and it generates computation task $T_{v}$ when it is sensing environment. We consider a new status update $i$  of the environment at the vehicle $v$ is sampled at time   $K^v_i$. The  sampling time is slotted with slot index $i \in \{1,2, \dots, K^V\}$.

\emph{Offloading and edge clouds:}  
Handling compute-intensive tasks in OBU may exhaust the computational resource and energy of the autonomous vehicle. In such a situation, the vehicle can offload computation tasks to edge clouds via communication links. We assume that every vehicle $v \in \mathcal{V}$  moves in the area covered by one or more RATs.  In our system model, 5G O-RAN uses  O-RAN Radio Units (O-RUs), while LTE uses Radio Unit (RU), evolved NodeB Distributed Unit (eNB-CU), and evolved NodeB Control Unit (eNB-CU). For  Wi-Fi,  we consider Wi-Fi hotspots with wired interface Y2 between Wi-Fi hotspot and N3IWF for the transport of traffic data and control data in 5G \cite{IEEE802}. We consider 5G, LTE, and Wi-Fi as a single multi-RAT network using ATSSS. Therefore, hereafter, unless stated otherwise, we use the terms ``RAT'' to mean ``RU'', or ``O-RU''  or ``Wi-Fi hotspot''. At least one RAT connection needs to be activated in the autonomous vehicle to offload data to proximity edge clouds. Let $\mathcal{S}$ denotes the set of RATs. Each vehicle $v$ is connected to a RAT $s \in \mathcal{S}$ via a wireless channel (W. Ch). Furthermore, we consider each RAT $s \in \mathcal{S}$ is connected to an Edge Cloud (EC) $r \in \mathcal{R}$ via fronthaul/Y2 link  of the capacity $\omega^s_{v, r}$. Let $\mathcal{R}$ be the set of ECs, where each EC $r$ accommodates O-Cloud that hosts  O-RAN Control Unit (O-CU),  O-RAN Distributed Unit (O-DU), eNB-DU, and eNB-CU. Using Multi-Access Edge Computing (MEC) server(s), each EC $r \in \mathcal{R}$ has computing resources $P_r$ that can be allocated to autonomous vehicles. Here, we consider MEC is collocated with UPF at EC. Each  EC $r \in \mathcal{R}$ serves multiple RATs and vehicles. When EC $r \in \mathcal{R}$ does not have enough resources, it can rather collaborate with nearby ECs than send tasks to the remote RC or data center. Unless stated otherwise, we use the terms “regional cloud” and “data center” interchangeably. Therefore, we define a CS of ECs in Section \ref{subsec:HLF}.  The collaboration of ECs helps in maximizing the utilization of edge resources and meeting computation deadlines. To enable such collaboration and application-level data exchange, we consider interface $Xn$ \cite{huang2020prospect} between O-RAN nodes (O-CU-CPs and O-CU-UPs) and interface $X2$ between eNB-CUs and O-CU-UPs.

\emph{Offloading and Regional Cloud (RC):} In the worst-case scenario, when resources are not available at any EC in the CS around a vehicle, the tasks of vehicles can be offloaded to the RC.  Each EC $r \in \mathcal{R}$ can access the RC via a wired backhaul of capacity $\omega_{r, RC}$.  We denote the computation capacity of the RC by $P_{RC}$.

\section{Task Offloading in Multi-RAT Edge Computing}
\label{sec:FederatedLearningl}
When a vehicle does not have enough computation and energy resources, its tasks can be offloaded to an EC. However, each EC has limited resources. Therefore, each EC needs to collaborate with other nearby ECs to process data at the edge of the network in proximity to autonomous vehicles. This section describes our Collaboration Space (CS) of ECs, communication model to reach ECs, and computation model. Unless stated otherwise, we use the terms “EC” and “edge server/MEC server” interchangeably.

\subsection{Collaboration Spaces of Edge Clouds}
\label{subsec:HLF}

We assume that ECs' network topology and locations are known at Non-RT RIC to form CSs. We assume that the ECs' network topology does not change frequently.  Given the locations of ECs, we propose Affinity Propagation-based  Algorithm for CS (APACS).  Affinity Propagation (AP) is a clustering algorithm based on messages passing between dataset elements to form clusters \cite{refianti2016performance}.  We choose AP over other approaches because AP is a fast clustering approach in terms of computation speed and does not depend on the initialization of the number of clusters, unlike most exiting clustering approaches such as k-means  \cite{refianti2016performance}. The APACS is implemented at Non-RT RIC, where the Non-RT RIC runs Algorithm \ref{algo:OKM} when the network topology changes. In the algorithm, we consider $z_r$ and $z_j$ to be the locations for ECs $r, j \in \mathcal{R}$ and $Z$ to be a function that computes the similarity between any two locations. When $Z(r, j) > Z(r, w)$,  $z_r$ is more similar or closer to $z_j$ than to $z_w$. Furthermore, we use squared distance of two locations for  $z_r$ and $z_w$ such that:
\begin{equation}
	Z(r, w) =-  \lVert z_r-z_w\rVert^2.
\end{equation}
When $Z(r, w)$ is large, we have high similarity. Furthermore, the APACS takes as input measurement of similarity between each pair of EC locations and exchanges messages between these locations until the locations of Centroid Edge Cloud (CECs) and corresponding locations of ECs gradually emerge as CSs.

In the messages exchange between ECs, we have responsibility and availability metrics. We use $\vect{\Phi}$  to denote the responsibility matrix, where  $\vect{\Phi}$ contains values $\phi(r, w)$ that shows how befitting $z_w$  as a location of CEC $w$ for $z_r$, by comparing it to other candidate locations of CECs for $z_r$.   The responsibility $\phi(r, w)$ is send from location $z_r$ of EC $r$  to candidate location $z_w$  of CEC $w $ as evidence of how well-suited EC $w \in \mathcal{R}$   is to serve as the CEC for EC $r \in \mathcal{R}$. The $\phi(r, w)$ takes into account other potential CECs for the EC $r$.  Furthermore,  we use $\vect{\Theta}$ to denote the availability matrix. The $\vect{\Theta}$ has values $\theta (r, w)$ that represents how appropriate it would be  $z_r$ to take $z_w$ as its CEC, taking into account other locations preference for $z_r$ as a CEC.  The availability $\theta (r, w)$ sent from CEC $w$ to EC $r$ reflects the accumulated evidence for how appropriate it would be for EC $r$ to choose EC $w$ as its CEC.  The $\vect{\Theta}$ considers other ECs that may select EC $w$  to be a CEC.

The proposed Algorithm \ref{algo:OKM} (APACS) starts by initializing 
the messages $\phi$  and $\theta$  to  zeroes.  Then, APACS needs to update $\phi$  and $\theta$  iteratively. The following equation updates for  responsibility  $\phi$ : 
\begin{equation}
	\begin{aligned}
		\phi(r, w)\leftarrow Z(r, w)-\underset{w \neq w'}{\text{max}}\{\; \theta (r, w') + Z(r, w')\}.
	\end{aligned}
	\label{eq:responsibility_matrix}
\end{equation}
In (\ref{eq:responsibility_matrix}), $\phi(r, w)$ uses the similarity between location $z_r$ of EC $r$ and location $z_w$  of CEC $w$ as input minus the largest of the similarities and availability between location $z_r$ and other candidate CEC  $w'$. Through iterations, when some locations are effectively assigned to other CECs, their availability $\theta$ will continue being reduced. Therefore, responsibility update enables all candidate CECs  to compete for owning ECs, where each CEC and its associated ECs form one CS. In other words, a CEC refers to an EC that is at the center of each CS, and the CEC is unique in each CS. Furthermore,  each EC belongs to one CS. The APACS updates availability  $\theta$ by using the following equation:
\begin{equation}
	\begin{aligned}
		\theta (r, w)\leftarrow \underset{r \neq w}{\text{min}}( \,0,\phi(w, w)+ \sum_{r' \notin (r,w)}\underset{}{\text{max}} (0,\phi(r', w) )) \,,
	\end{aligned}
	\label{eq: availability_matrix1}
\end{equation}
\begin{equation}
	\begin{aligned}
		\theta (w, w)\leftarrow \sum_{r' \neq w} \underset{}{\text{max}}(0,\phi(r', w)).
	\end{aligned}
	\label{eq:availability_matrix2}
\end{equation}
In (\ref{eq: availability_matrix1}), we consider the availability $\theta (r, w)$ is equal to the self-responsibility $\phi(w, w)$ plus the sum of the positive responsibilities that the candidate CEC $w$ receives from other ECs. In (\ref{eq: availability_matrix1}), we consider only positive responsibilities; thus, a good CEC is the one that has higher similarities. Furthermore, in (\ref{eq:availability_matrix2}), we consider $\phi$  and $\theta$  can be combined at any stage to decide the number of ECs that are CECs. In other words, the number of CECs equals the number of CSs, i.e., the number of clusters. 

To prevent  Algorithm \ref{algo:OKM}  to run indefinitely, for each location $z_r$, we choose the value of $z_w$ that maximizes the following criterion:
\begin{equation}
	c(r,w)=	\phi(r,w) + \theta(r,w).
\end{equation}
In the criterion, $r$ is the row and $w$ is the column of the associated matrix of responsibility and availability. Each EC with the highest criterion value at each row is the CEC. Furthermore, rows that share the same CEC are in the same CS.  Algorithm \ref{algo:OKM}  performs iterations until either the CS boundaries remain unchanged or the algorithm reaches the maximum number of iterations $b_m$.
\begin{algorithm}[t]
	\caption{: AP-based  Algorithm for CS (APACS).}
	\label{algo:OKM}
	\begin{algorithmic}[1]
		\STATE{\textbf{Input:} $\mathcal{R}$: A set of ECs with their coordinates, $b_m$: Maximum number of iterations; }
		\STATE{\textbf{Output:} $\vect{\Theta}$: Availability matrix,  $\vect{\Phi}$: Responsibility matrix, and  number of CSs;}
		\STATE{Initialize iteration $b_i=0$, $\vect{\Theta}\leftarrow\emptyset$, and $\vect{\Phi}\leftarrow\emptyset $;}
		\FOR{EC  $r \in \mathcal{R}$ and $b_i\leq b_m$}                    
		\STATE {Select EC  $w$;}
		\STATE{Compute similarity $Z(r, w)$ between any two locations of EC $r$ and $w$;}
		\STATE{Compute responsibility $\phi$ using (\ref{eq:responsibility_matrix});}
		\STATE{Compute availability $\theta$ using (\ref{eq: availability_matrix1}) and (\ref{eq:availability_matrix2});}
		\STATE {Use $\theta$ and $\phi$ to compute criterion $c(r,w)$;}
		\STATE{$\vect{\Theta} \leftarrow \theta$ and $\vect{\Phi} \leftarrow \phi$;}
		\STATE{Find maximum $c(r,w)$ for each EC $r$ and $w$, find CECs and associate ECs to CECs for forming CSs;}
		\STATE{$b_i=b_i+1$;}	
		\STATE{Return to step $4$;}		
		\ENDFOR
		\STATE{Via  Near-RT RIC, Non-RT RIC informs ECs about their CSs.}
	\end{algorithmic}
\end{algorithm}

The APACS makes non-overlapping CSs of ECs.  Since the CEC is the center of each CS, to facilitate the communication between ECs of different CSs, we assume that the CECs can exchange information.  The information includes location and available resources in the CS.  However, to avoid overhead due to information exchange between CSs, one-hop distance can be applied. When there is no available resource in its CS, an EC can redirect the task to another EC of another CS or to the RC. Such intercluster routing is defined in \cite{lin2020energy}.  In this work, we focus on intra-cooperation between ECs that belong to the same CS. We consider inter-cooperation between ECs that belong to different CSs as future work. Within a CS, ECs exchange resource utilization information such as CPU and memory. Each EC stores this information in the resource allocation table defined in \cite{ndikumana2019intelligentedge}.
\subsection{Communication Model for Autonomous Vehicles}
\label{subsec:communication model}   
\begin{figure}[t]
	\centering
	\includegraphics[width=1.0\columnwidth]{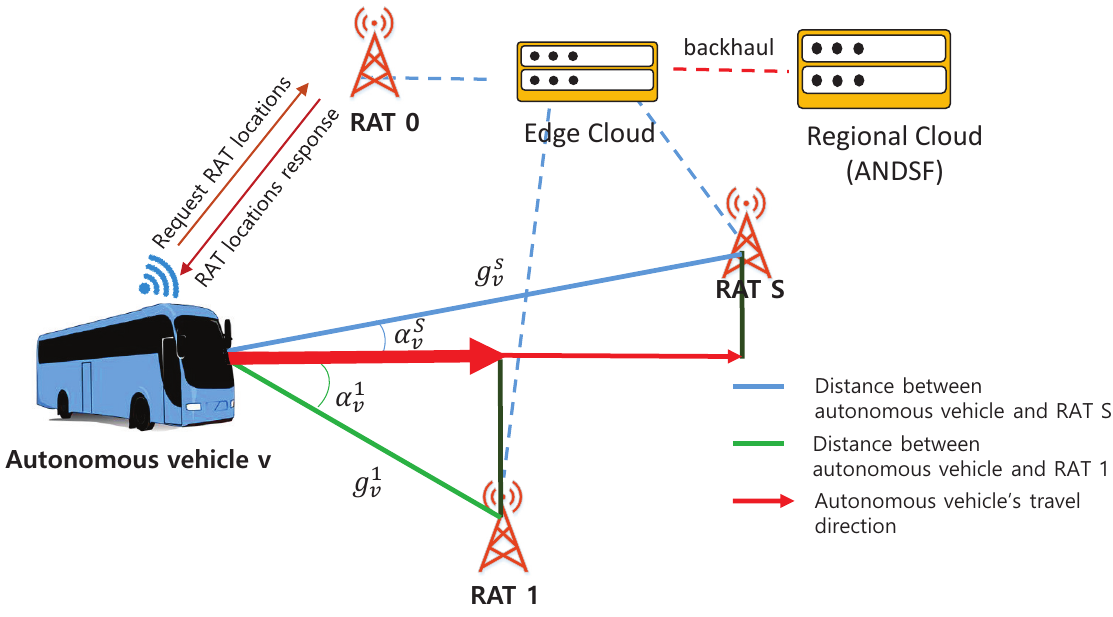}
	\caption{Communication planning model.}
	\label{fig:communication_model2}
\end{figure}

Fig. \ref{fig:communication_model2} shows the communication planning model. In the model, before the autonomous vehicle starts its road trip, it preselects RATs available in its route for offloading tasks when its resource  exhausted. We assume each vehicle $v \in \mathcal{V}$ moves in an area covered by RATs. To obtain RATs information such as coordinate and coverage, we assume that Access Network Discovery and Selection Function (ANDSF)  server \cite{3GPPTS} is available at RC to enable network discovery and selection between 3GPP and non-3GPP access networks. The vehicle sends a request to the ANDSF using its home RAT denoted RAT $0$ in Fig. \ref{fig:communication_model2}. The request includes the geographic location and destination of the vehicle. On the other hand, the ANDSF’s feedback contains the coordinates and coverage areas of all available RATs along the vehicle trajectory.  Then, each vehicle $v$ calculates the distance $\tilde{d}^s_v$ between its route and each RAT $s$:
\begin{equation}
	\label{eq:O-RU-road}
	\begin{aligned}
		\tilde{d}^s_v= g^s_v sin \alpha^s_v,
	\end{aligned}
\end{equation}
where $\alpha^s_v$ is the angle between the trajectory of movement of autonomous vehicle $v$ and the line from RAT $s \in \mathcal{S}$. We denote by $g^s_v$ a geographical distance between vehicle $v$ and RAT $s$. Also, each vehicle $v$ calculates the remaining distance $d^v_s$ to reach each area covered by  RAT $s$:
\begin{equation}
	\label{eq:distancevehicle_O-RU}
	\begin{aligned}
		d^v_s= g^s_v cos \alpha^s_v.
	\end{aligned}
\end{equation}
Then, the autonomous vehicle computes the probability  $\chi^s_v$ that RAT $s$ is preselected for being used to offload computation task $T_{v}$ to EC  such that:
\begin{equation}
	\setlength{\jot}{10pt}
	\chi^s_v=
	\begin{cases}
		1,\; \text{if $\tilde{d}^s_v=0$}, \\
		\frac{\tilde{d}^s_v}{\gamma_s}\;\text{if $0<\tilde{d}^s_v<\gamma_s$,}\\
		0,\;\text{otherwise,}
		\label{eq:probability_O-RU}
	\end{cases}
\end{equation}
where $\gamma_s$ is the area covered by  RAT $s$. Furthermore, based on a speed $\iota_v$  of  vehicle $v$, we define $t^s_{v}$ as the time required by  vehicle $v \in \mathcal{V}$ to leave each area covered by RAT $s$. We can calculate $t^s_{v}$ as follows:
\begin{equation}
	\setlength{\jot}{10pt}
	t^s_{v}=\frac{ \gamma_s}{\iota_v}. 
\end{equation}
Once the vehicle reaches an area $\gamma_s$, it can select RAT $s$ among preselected multiple RATs.  When $t^s_{v} \leq \tilde{\tau}_{v}$, the autonomous vehicle can easily offload the computation task and get output in the area covered by RAT $s$. However, when $\tau^s_{v}> \tilde{\tau}_{v}$, the autonomous vehicle can select the next RAT to use for offloading computation task to EC.

When RAT $s$ is Wi-Fi, we consider the Wi-Fi channel is shared to vehicles via a contention-based model as described in \cite{cheng2016opportunistic}. Therefore, the instantaneous data rate for vehicle $v$ via Wi-Fi is given by:
\begin{equation}
	\rho^{s,w}_v=\frac{\varphi_s \rho^{s} \xi^s_v(|\mathcal{V}_s|)}{|\mathcal{V}_s|}, \forall v \in \mathcal{V}_s,\; s \in \mathcal{S}, 
	\label{eq:instantaneous_data_WAP}
\end{equation} 
where $\varphi_s $ is the Wi-Fi throughput efficiency factor. The  $\varphi_s $ is used to determine overhead related to MAC protocol layering such as header, Distributed Coordination Function Interframe Space (DIFS), Short Interframe Space (SIFS), and acknowledgment (ACK). $|\mathcal{V}_s|$ is the number of vehicles that communicated  simultaneously with Wi-Fi $s$, where $\mathcal{V}_s \subset \mathcal{V}$. Furthermore, $\rho^{s}$ is the maximum theoretical data rate that Wi-Fi can handle \cite{cheng2016opportunistic}.  In (\ref{eq:instantaneous_data_WAP}), we denote by $\xi^s_v(|\mathcal{V}_s|)$  as decreasing function, which is a function of the number of autonomous vehicles connected to Wi-Fi. $\xi^s_v(|\mathcal{V}_s|)$ helps determine the impact of contention over Wi-Fi throughout.

When RAT $s$ is cellular, we consider orthogonal resource allocation. We assume that each vehicle can offload its  task when there is enough spectrum resource to satisfy its task offloading. Therefore, the spectrum efficiency for autonomous vehicles $v$ becomes:
\begin{equation}
	\label{eq:SINR}
	\begin{aligned}
		\varrho^s_v= \log_2\left(1 + \frac{\varkappa_v |G^s_v|^2}{\sigma_v^2}\right),  \;\forall v \in \mathcal{V},\; s \in \mathcal{S},
	\end{aligned}
\end{equation}
where $\varkappa_v$ is the transmission power of autonomous vehicle $v$ and $G^s_v$ is the channel gain between vehicle $v$ and the O-RU/RU $s$.  We define $\sigma_v^2$ as the power of the Gaussian noise at vehicle $v$. The achievable data rate of the  vehicle for offloading its computational task via O-RU/RU $s$ is given by:
\begin{equation}
	\rho^{s,c}_v= a_v^s \omega_{v, s} \varrho^s_v, \forall v \in \mathcal{V},\; s \in \mathcal{S}.
	\label{eq:instantaneous_data}
\end{equation}
Each vehicle obtains a fraction $ a_v^s$ of bandwidth capacity $\omega_{v, s}$ such that $\sum_{v\in \mathcal{V}_r} a^s_{v}=1$. 

We consider the vehicle $v$ can perform handover between Wi-Fi and cellular network. Since the strength of a wireless signal gets attenuated with distance, the vehicle needs to select cellular or Wi-Fi based on achievable data rate and the time required to leave the area covered by RAT. Therefore, we define connection variables $\eta_v^{s,w}$ for Wi-Fi and  $\eta_v^{s,c}$ for cellular network:
\begin{equation}
	\setlength{\jot}{10pt}
	\eta_v^{s,w} =
	\begin{cases}
		1,\; \text {if $d^v_s=0$, 
			$\rho^{s,w}_v> \rho^{s,c}_v$ $\chi^s_v > 0$, and  $t^s_{v} \leq \tilde{\tau}_{v}$}, \\
		0,\;\text{otherwise,}
	\end{cases}
\end{equation}
\begin{equation}
	\setlength{\jot}{10pt}
	\eta_v^{s,c} =
	\begin{cases}
		1,\; \text {if $d^v_s=0$, 
			$\rho^{s,c}_v \geq \rho^{s,w}_v$ $\chi^s_v > 0$, and  $t^s_{v} \leq \tilde{\tau}_{v}$}, \\
		0,\;\text{otherwise.}
	\end{cases}
\end{equation}
Furthermore, we define $x_v^{s\rightarrow r}$ as a decision variable that indicates whether or not the autonomous vehicle uses RAT $s \in  \mathcal{S}$  to offload its task to EC $r$: 
\begin{equation}
	\setlength{\jot}{10pt}
	x_v^{s\rightarrow r} =
	\begin{cases}
		1,\; \text{$T_v$ is offloaded from vehicle $v$ to EC $r$} \\  \; \; \;\;  \;\text {via RAT $s$ if $\eta_v^{s,w} + 	\eta_v^{s,c}=1$}, \\
		0,\;\text{otherwise.}
		\label{eq:probability_O-RU_variable}
	\end{cases}
\end{equation}
Equations (\ref{eq:probability_O-RU}) and (\ref{eq:probability_O-RU_variable}) guarantee that the vehicle has an active RAT connection. Once vehicle $v$  reaches an area covered by  RAT $s \in \mathcal{S}$, it can immediately start offloading its computation task, i.e., when its local computation resource is not enough.

We model the network linking for RAT, edge, and regional clouds as an undirected graph $\mathcal{G}=(\mathcal{O},\mathcal{E})$, where $\mathcal{O}$ denotes the set of routers/switches and $\mathcal{E}$ is the set of links. 	Furthermore,  we denote by $\mathcal{A}\subset \mathcal{E} $  the set of fronthaul/Y2 links, by $\mathcal{B}\subset \mathcal{E}$ the set of the links between ECs, and by $\mathcal{C}\subset \mathcal{E}$ the set of the backhaul links, where $ \mathcal{E}=\mathcal{A}\cup \mathcal{B} \cup \mathcal{C}$. %We assume all  O-CUs and O-DUs are hosted in O-Clouds. 
For each fronthaul/Y2 link $a\in \mathcal{A}$, the traffic volume is expressed as:
\begin{equation}
	\rho_{A}= \sum \nolimits_{v \in \mathcal{V}(a)} x_v^{s\rightarrow r} s{d_v},
	\label{eq:interalloadFronthaul}
\end{equation}
where $\mathcal{V}(a)\subset \mathcal{V}$ is a set of  vehicles that use the fronthaul/Y2 link $a\in \mathcal{A}$. We assume that all offloading tasks reach ECs via RATs using fronthaul/Y2 links. Futhermore, for each link $b\in \mathcal{B}$ between ECs, the traffic volume is expressed as:
\begin{equation}
	\rho_{B}= \sum \nolimits_{v \in \mathcal{V}(b)} (1-y_v^{s\rightarrow r}) x^r_v s{d_v},
	\label{eq:interalloadedge}
\end{equation}
where $\mathcal{V}(b) \subset \mathcal{V}$ is a set of vehicles  using the link $b$. Here, $y_v^{s\rightarrow r}$ is a decision variable that indicates whether or not the task of autonomous vehicle $v$ offloaded via RAT $s$ is computed at EC $r$.
\begin{equation}
	\setlength{\jot}{10pt}
	y_v^{s\rightarrow r} =
	\begin{cases}
		1,\; \text{if  task $T_v$  of vehicle $v$ offloaded }
		\\ \; \; \; \;\text{via RAT $s$ is computed at EC $r$,}\\
		0,\;\text{otherwise.}
	\end{cases}
\end{equation}

In a CS, when $y_v^{s\rightarrow r}=1$, the EC $r$ computes the task $T_v$ and does not forward task $T_v$ to another EC. However, when EC $r$ does not have enough resources, it can offload the task to another EC $j$ of the same CS, which has enough resources. Therefore, we define  $y_v^{r\rightarrow j}$ as a decision variable that indicates whether or not the task of vehicle $v$ offloaded to EC $r$  is redirected to EC $j$ for computation.  
\begin{equation}
	\setlength{\jot}{10pt}
	y_v^{r\rightarrow j} =
	\begin{cases}
		1,\; \text{if task $T_v$ offloaded to EC $r$ is }
		\\ \; \; \; \;\text{redirected to EC $j$ for computation,}\\
		0,\;\text{otherwise.}
	\end{cases}
\end{equation}

In the worst-case scenario, when the resources are not enough at ECs, the task can be offloaded to regional cloud. Therefore, for each backhaul link $c \in \mathcal{C}$, the traffic volume is given by:
\begin{equation}
	\rho_{C}= \sum \nolimits_{v \in \mathcal{V}(c)} (1-(y_v^{s\rightarrow r}+y^{r\rightarrow j}_v)) x^r_v s{d_v},
	\label{eq:interalload}
\end{equation}
where $\mathcal{V}(c) \subset \mathcal{V}$ is a set of autonomous vehicles using the backhaul link $c$. When a task $T_v$ is computed in CS, i.e., at EC $r$ ($y_v^{s\rightarrow r}=1$)  or EC $j$ ($y^{r\rightarrow j}_v=1$), the  task   will  not be offloaded to regional cloud.

\subsection{Computation Model}
\label{subsec:CollobaorationSpace}
\begin{figure}[t]
	\centering
	\includegraphics[width=1.0\columnwidth]{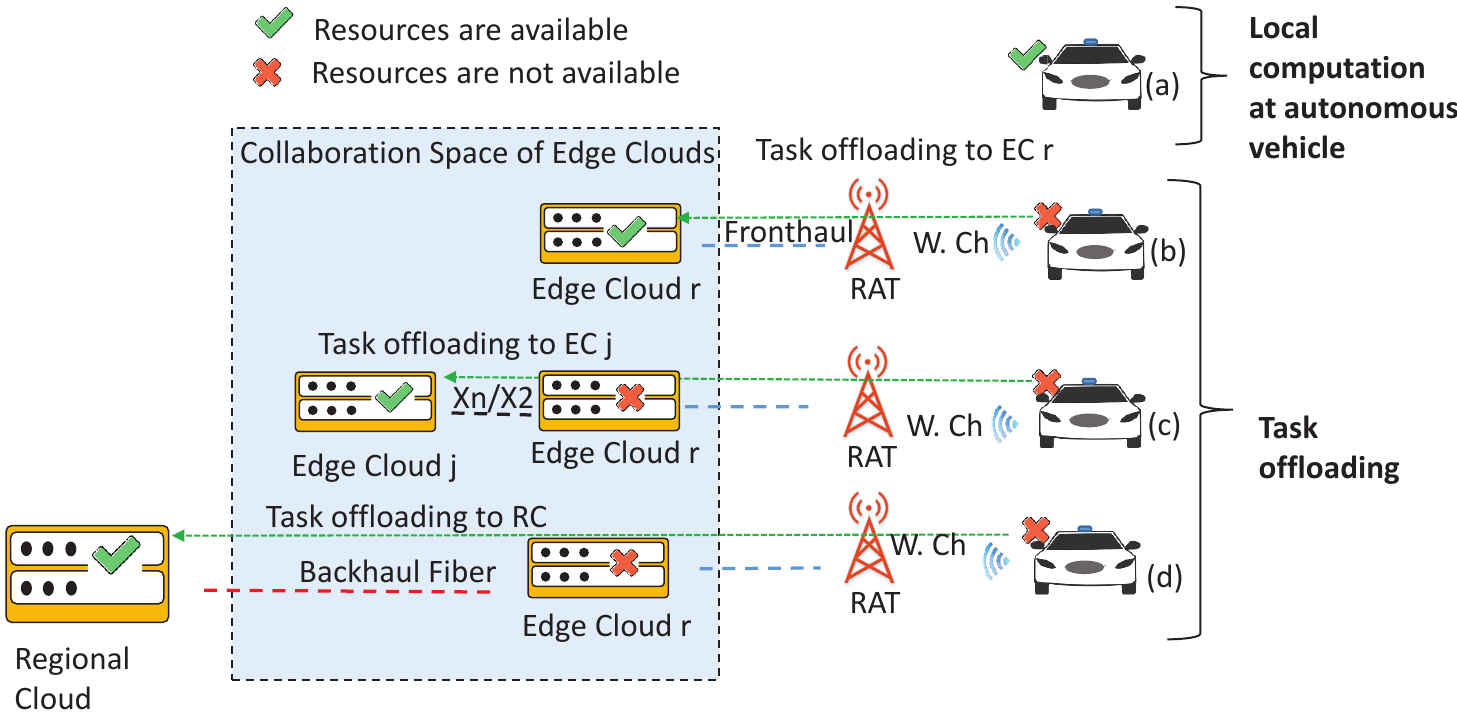}
	\caption{Task offloading using communication model.}
	\label{fig:SystemModel2}
\end{figure}

Fig. \ref{fig:SystemModel2} shows task offloading using the proposed communication model and CS. In this subsection, we describe all computation scenarios presented in Fig. \ref{fig:SystemModel2}. 

%\subsubsection{ Handling Tasks Locally at Autonomous Vehicle}
%\label{subsec:TasksLocally}

\emph{Scenario (a):} Computing  task $T_{v}$  at autonomous vehicle $v \in \mathcal{V}$ requires CPU energy $E_{v}=  s{d_v}\nu \tilde{z}_{v}P_{v}^2$,
where $\nu$ is a hardware architecture's constant parameter. The computation of task $T_{v}$ takes execution time $\tau_v$, where $\tau_v$ is given by:
\begin{equation}
	\label{eq:compution_time}
	\tau_v=\frac{s{d_v}\tilde{z}_{v} }{P_{v}}.
\end{equation}
However, when $\tau_v> \tilde{\tau}_{v}$, or $ \tilde{z}_{v}>P_{v}$, or $E_{v}>\tilde{E_{v}}$ vehicle $v$ does not have enough resources to meet the computation deadline. Here, we consider $\tilde{E_{v}}$ as available energy for autonomous vehicle $v\in \mathcal{V}$. Therefore, we define vehicle status parameter $\alpha_{v}\in \{0,1\}$  for computing task $T_{v}$, where
\begin{equation}
	\setlength{\jot}{10pt}
	\alpha_{v} =
	\begin{cases}
		0,\; \text{if  $\tilde{z}_{v}>P_{v}$, or  $\tau_v>\tilde{\tau}_{v}$, or $ E_{v}>\tilde{E_{v}}$},\\
		1, \;\text{otherwise.}
	\end{cases}
\end{equation} 
We define the total local execution time $\tau^\textrm{loc}_{v}$ of task $T_{v}$ at vehicle $v$ to be:
\begin{equation}
	\setlength{\jot}{10pt}
	\tau^\textrm{loc}_{v} =
	\begin{cases}
		\tau_v\;, \text{if $\alpha_{v}=1$ and  $x_v^{s\rightarrow r}=0$},\\
		\tau_v+ \varphi_v\;, \text{if $\alpha_{v}=0$ and  $x_v^{s\rightarrow r}=0$},\\
		0, \; \text{if $\alpha_{v}=0$ and  $x_v^{s\rightarrow r}=1$},\\
	\end{cases}
\end{equation}
where $\varphi_v$ is the average waiting time of task $T_{v}$ for being executed at vehicle $v$ when the resources become available. If
the  task $T_{v}$  is computed locally, then the computation time  of update $i$ is $L^\textrm{loc}_{iv}=\tau^\textrm{loc}_{v}$.

\emph{Scenario (b):} When a vehicle $v$ does not have enough computation resources and it cannot wait until the resources become available, the vehicle can offload the task to its EC $r  \in \mathcal{R}$.
In considering wireless and fronthaul/Y2 links, the transmission delay for offloading task $T_{v}$ is:
\begin{equation}
	\tau_v^{s \rightarrow r}=\sum_{v \in \mathcal{V}_r}x_v^{s\rightarrow r}\left(\frac{ s{d_v}}{\eta_v^{s,w}	\rho^{s,w}_v + 	\eta_v^{s,c}\rho^{s,c}_v} + \frac{ s{d_v}}{\omega^s_{v, r}}\right), \forall s \in \mathcal{S}.
\end{equation}

The computation allocation $p_{vr}$ of the offloaded task $T_{v}$ from autonomous vehicle $v$ at  EC $r$ can be defined as follows:
\begin{equation}
	p_{vr}=P_r\frac{\tilde{z}_{v}}{\sum_{g \in \mathcal{V}_r}\tilde{z}_{g}},\; \forall v  \in \mathcal{V}_r,\; r \in \mathcal{R},
	\label{eq:ComputationAk}
\end{equation}
where $\sum_{g \in \mathcal{V}_r}\tilde{z}_{g}$ is the computation tasks of other autonomous vehicles than $v$. 
Furthermore, at each  EC $r$, the total computation allocations must satisfy the following computation constraint:
\begin{equation}
	\sum_{v\in \mathcal{V}_r}x_v^{s\rightarrow r}p_{vr}y_v^{s\rightarrow r}\leq P_r,\; \forall  r \in \mathcal{R}.
\end{equation}
For offloaded task  $T_{v}$ at EC $r$, the execution latency $\tau_{vr}=\frac{s{d_v}\tilde{z}_{v} }{p_{vr}}$. Therefore, the total execution time for task $T_v$ at  EC $r$ becomes:
\begin{equation}
	\tau^e_{vr}= \tau^{s\rightarrow  r}_v + \tau_{vr},\; \forall v \in \mathcal{V}_r, \;s \in \mathcal{S}, \; r \in \mathcal{R}.
\end{equation}
However, the EC $r$ may be overloaded while still having resources ($\tilde{z}_{v}\leq p_{vr} $) to meet computation deadline $\tilde{\tau}_{v}$ of vehicle $v$ ($\tau^e_{vr} \leq \tilde{\tau}_{v}$). When $\tilde{\tau}_{v}$ is large enough, EC $r$ can forward task $T_{v}$ to another  EC $j$ having a lower computational load. This requires analyzing computation load in both ECs $r$ and $j$ to ensure that the task forwarding will not perturb the vehicle service. Also, to minimize propagation delay, EC $j$ should be in less distance than RC.

\emph{Scenario (c):} When  $\tilde{z}_{v}>p_{vr} $ or $\tau^e_{vr} > \tilde{\tau}_{v}$, i.e.,  EC $r$ does not have enough resources to meet  computation deadline.
The EC $r$ checks its resource allocation table and find EC $j$  which is in less distance than RC and has enough resources to compute  $T_{v}$. Then, EC $r$ offloads the task to the EC $j$. The computation resource allocation $p_{vj}$ at EC $j$ can be calculated using similar approach in (\ref{eq:ComputationAk}). Using $p_{vj}$,  the execution latency
$\tau_{vj}=\frac{s{d_v}\tilde{z}_{v} }{p_{vj}}$  for task  $T_{v}$ at EC $j$. Therefore, the total execution time for a  task offloaded by autonomous vehicle $v$ to  EC $j$ becomes:
\begin{equation}
	\tau^e_{vrj}=\tau_v^{s \rightarrow r} + \tau^{r\rightarrow j}_v+ \tau^{r\rightarrow j}+ \tau_{vj},\; \forall v \in \mathcal{V}_r, \text{ and } r,j \in \mathcal{R},
\end{equation}
where $\tau^{r\rightarrow j}_v=\frac{\sum_{v \in \mathcal{V}_r}y^{r\rightarrow j}_v s{d_v}}{\omega^j_r}$ is the offloading delay between  EC $r$ and  EC $j$. Here,    $\omega^j_r$ is the link capacity between  EC $r$ and  EC $j$. Furthermore, we denote  $\tau^{r\rightarrow j}=\frac{h^{r\rightarrow j}}{\kappa^{r\rightarrow j}},\; \forall r,j \in \mathcal{R}$ as the propagation delay between EC $j$ and EC $r$. We use $h^{r\rightarrow j}$ to represent the length of physical link between EC $j$ and EC $r$ and $\kappa^{r\rightarrow j}$ to represent propagation speed. 

\emph{Scenario (d):} In the worst-case scenario, there is no available resource in the CS. In other words, EC $r$ does not have enough resources, and there is no other EC $j$, which is in less distance than RC and has computation resources to handle the task $T_{v}$.  Therefore, we define $y^{r\rightarrow RC}_v$ as a  computation decision variable, where $y^{r\rightarrow RC}_v$ is expressed as follows:
\begin{equation}
	\setlength{\jot}{10pt}
	y^{r\rightarrow RC}_v =
	\begin{cases}
		1,\; \text{if  task $T_v$  of vehicle $v$ is offloaded to RC by}
		\\ \; \; \; \;\text{EC $r$ and  $\rho_{B} \leq \omega_{r, RC}$,}\\
		0,\;\text{otherwise.}
	\end{cases}
\end{equation}
Furthermore, we define $\tau^{r\rightarrow RC}_v=\frac{\sum_{v \in \mathcal{V}_r}y^{r\rightarrow RC}_vs{d_v}}{\omega_{r, RC}}$ as the offloading delay between EC $r$ and RC, where $\omega_{r, RC}$ is the link capacity between  EC $r$ and remote RC. Therefore, the total execution time for task $T_v$  offloaded by autonomous vehicle $v$ at RC becomes:
\begin{equation}
	\tau^e_{vrRC}=\tau_v^{s \rightarrow r} + \tau^{r\rightarrow RC}_v + \tau^{r\rightarrow RC} + \tau_{vRC},\; \forall v \in \mathcal{V}_r, \text{ and } r \in \mathcal{R},
\end{equation}
where  $\tau^{r\rightarrow RC}=\frac{h^{r\rightarrow {RC}}}{\kappa^{r\rightarrow {RC}}}$ is the propagation delay between EC $r$ and RC. Here, $h^{r\rightarrow {RC}}$ is  the length of physical link between EC $r$ and RC, and $\kappa^{r\rightarrow {RC}}$ is propagation speed. Furthermore, at RC, execution latency $\tau_{vRC}=\frac{s{d_v}\tilde{z}_{v} }{P_{RC}}$, where $P_{RC}$ is computation resource allocation.

\section{Problem Formulation and Solution} 
This section discusses the problem formulation, the proposed solution, and the application scenario of the AoP-based offloading.
\label{sec:Problem_Formulation}
\subsection{Problem Formulation}
\label{subsec:ProblemFormulationn}
Considering all scenarios $(a)$, $(b)$, $(c)$ and $(d)$, the total offloading delay $\tau^\textrm{off}_{v}$ of task $T_v$   from autonomous vehicle $v$ is given by:
\begin{equation}
	\begin{aligned}
		\tau^\textrm{off}_{v}=y_v^{s\rightarrow r}\tau^e_{vr} + y^{r\rightarrow j}_v\tau^e_{vrj}+ y^{r\rightarrow RC}_v\tau^e_{vrRC}.
	\end{aligned} 
\end{equation}
%When the  vehicle $v$ computes its task $T_v$ locally, the computational delay of $\tau^\textrm{loc}_v$ is required. 
When vehicle $v$  decides to offload its computational task to EC or RC, offloading and computation delays $\tau^\textrm{off}_{v}$ are required. In other words, when the  task $T_{v}$  is computed at EC $r$ or RC, then the computation time of update $i$ is $L^\textrm{off}_{iv}=\tau^\textrm{off}_{v}$. Therefore, we consider total  offloading and computation time:
\begin{equation}
	\begin{aligned}
		&L^v_i= (1-x_v^{s\rightarrow r})L^\textrm{loc}_{iv}  +x_v^{s\rightarrow r}(L^\textrm{off}_{iv}).
	\end{aligned}
	\label{eq:total_compution_cost}
\end{equation}
After computation, we consider $M^v_i$ as the time to deliver the processed result of update $i$ at autonomous vehicle $v$.
\begin{equation}
	\begin{aligned}
		& M^v_i = K^v_i +L^v_i.
	\end{aligned}
	\label{eq:get_results}
\end{equation} 
In other words, $M^v_i$ is the time vehicle $v$ receives the processed result of update $i$.  

We consider the autonomous vehicle can generate a new task after time $N^v_i \geq 0$. Therefore, sampling the new status update $i+1$ is done at time $k^v_{i+1} = M^v_i+N^v_i$. At any time $t$, the freshest status update $i$ at the vehicle $v$ becomes:
\begin{align}
	u_v(t)=\max \{K^v_i : M^v_i \leq  t; \forall i \}
	\label{eq:age_of}.
\end{align}
Therefore, at time $t$, let us denote
$a^v(t)$ as instantaneous AoP of autonomous vehicle $v$, where  $a^v$ is given by:
\begin{equation}
	a^v(t) = t-u_v(t).
\end{equation}
Therefore, the overall AoP  $A^v$ of autonomous vehicle $v$ becomes:
\begin{equation}
	\label{eq:proce_time}
	A^v=\lim_{t\to\infty} \frac{1}{t} \int_{0}^{t} a^v(t)dt.
\end{equation}
AoP, which represents data freshness, is crucial for autonomous driving. The status information of the environment impacts the future behavior of autonomous driving, i.e., the time of sampling the new status update of the environment. In other words, AoP captures the time elapsed from status being generated at the vehicle to the latest status update after computation. AoP and network delay represent different metrics. Network delay is a packet-based performance metric, expressing the time elapsed between the packet generation at the source and reception at the destination without considering computation delay. Also, computation delay or execution delay does not include network delay. Therefore, we choose AoP over other metrics because it includes network delay and computation delay that affect the time of sampling a new status update.

\begin{figure}[t]
	\centering
	\includegraphics[width=0.95\columnwidth]{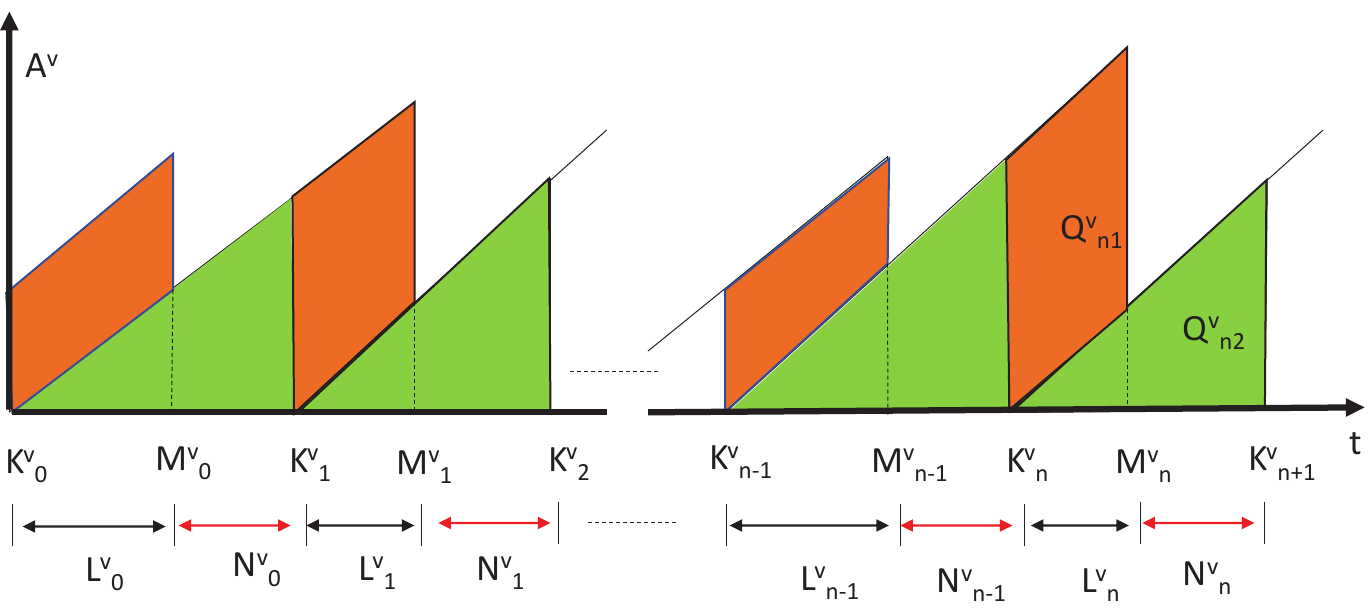}
	\caption{Age-based task offloading.}
	\label{AoP_offlaoding}
\end{figure}

To compute the average AoP, as described in \cite{li2021age},   we can decompose integral using a series of areas in Fig. \ref{AoP_offlaoding}.  The shaded parallelogram $Q^v_{i1}$ defines:
\begin{equation}
	\label{eq:Q1}
	Q^v_{i1}=(L^v_{i-1} + N^v_{i-1})L^v_i,
\end{equation}
and shaded triangle $Q^v_{i2}$ defines:
\begin{equation}
	\label{eq:Q2}
	Q^v_{i2}=\frac{1}{2}(L^v_{i} + N^v_{i})^2.
\end{equation}
Therefore, the average AoP $A^{av}$ becomes:
\begin{equation}
	\label{eq:averageAoP1}
	A^{av}=\frac{\sum_{i\rightarrow \infty}Q^v_{i1}+Q^v_{i2}}{\sum_{i\rightarrow \infty}L^v_{i} + N^v_{i}}.
\end{equation}
At time $M^v_i$, we denote $\Omega^v_i \triangleq \{ L^v_{i-1},  N^v_{i}, L^v_{i}\}$ as
the  system state, where $\Omega$ is the system state space. We consider the system state $\Omega$ to be finite, where $|\Omega| = K^V$. Therefore,  the autonomous vehicle can choose an action $\digamma^v_i \triangleq \{x_v^{s\rightarrow r}, y_v^{s\rightarrow r}, y^{r\rightarrow j}_v, y^{r\rightarrow RC}_v\}$ from the action space $\digamma$. The action space $\digamma$ consists  of offloading and computation decisions. Therefore, we need  offloading and computation policy $\pi$. The policy $\pi$  is defined as a mapping from the system state space $\Omega$ to the  action space $\digamma$, where $\pi: \Omega \rightarrow \digamma$. Let us consider $Q^v_{i}=Q^v_{i1}+Q^v_{i2}$.  When a policy $\pi$ is employed, the average AoP  can be computed as follows:
\begin{equation}
	\label{eq:averageAoP}
	A^{av}(\vect{\pi})=\underset{n\rightarrow \infty}{\text{lim sup}}\frac{\E_{\pi}[\sum_{i=1}^{n}Q^v_{i}]}{\E_{\pi}[\sum_{i=1}^{n}L^v_{i} + N^v_{i}]}.
\end{equation}

We aim to find the optimal policy $\vect{\pi}^*$ that minimizes the average AoP as follows:
\begin{subequations}\label{eq:problem_formulation}
	\begin{align}
		&\underset{\pi}{\text{minimize}}\ \  \sum_{r\in \mathcal{R}}\sum_{v\in \mathcal{V}_r}	A^{av}(\vect{\pi})
		\tag{\ref{eq:problem_formulation}}\\
		&\text{subject to:}\nonumber\\
		& \sum_{v\in \mathcal{V}_r}x_v^{s\rightarrow r} a^s_{v}\leq 1, \;  \forall s \in \mathcal{S},\label{first:a}\\
		&\sum_{v\in \mathcal{V}_r}x_v^{s\rightarrow r}p_{vr}y_v^{s\rightarrow r}\leq P_r,\; \forall  r \in \mathcal{R}. \label{first:b}
	\end{align}
\end{subequations}
Constraint  (\ref{first:a}) ensures that a fraction of communication resource allocated to each autonomous vehicle $v$ must not exceed available communication resources. Constraint (\ref{first:b}) ensures that the computation resources allocated to autonomous vehicles do not exceed available computation resources. 

\subsection{Proposed Solution}
\label{sec:Proposed_Solution}

The formulated problem in  (\ref{eq:problem_formulation}) is computationally intractable to find the optimal policy $\vect{\pi}^*$. Therefore, to simplify the problem in (\ref{eq:problem_formulation}),  we formulate a surrogate function of the original problem (\ref{eq:problem_formulation}), where  a surrogate function  is defined as follows:
\begin{equation}
	\label{eq:surrogate}
	\tilde{A}^v(\vect{\pi})=\underset{n\rightarrow \infty}{\text{lim sup}}\frac{1}{n}\E_{\pi}[\sum_{i=1}^{n}Q^v_{i}]. 
\end{equation}	
Then, we minimize surrogate function of the original problem as follows:
\begin{equation}
	\begin{aligned}
		\label{eq:averageAosurrogate}
		&\vect{\pi}^*=\underset{\vect{\vect{\pi}}}{\text{minimize}}\ \  \sum_{r\in \mathcal{R}}\sum_{v\in \mathcal{V}_r}	\tilde{A}^v(\vect{\pi})\\
		&\ \text{subject to: (\ref{first:a}) and (\ref{first:b})}.
	\end{aligned}	
\end{equation}

To solve (\ref{eq:averageAosurrogate}), we  transform the formulated problem in (\ref{eq:averageAosurrogate}) to  unconstrained optimization problem by using Lagrangian method \cite{boyd2011distributed}. Therefore, the problem  (\ref{eq:averageAosurrogate}) becomes:
\begin{equation}
	\begin{aligned}
		\mathcal{L(\vect{\pi},\vect{\lambda}, \vect{\mu}})= \sum_{r\in \mathcal{R}}\sum_{v\in \mathcal{V}_r}	\tilde{A}^v(\vect{\pi}) +  \sum_{v\in \mathcal{V}_r}\lambda_v(x_v^{s\rightarrow r} a^s_{v} - 1) + \\ \sum_{v\in \mathcal{V}_r}\mu_v (x_v^{s\rightarrow r}p_{vr}y_v^{s\rightarrow r} - P_r),
	\end{aligned}
	\label{eq:problem_formulationLagrangianf}
\end{equation}
where $\lambda_v$ is the Lagrangian multiplier associated with the  constraint in (\ref{first:a}), while  $\mu_v$  is the Lagrangian multiplier
associated with the  constraint in (\ref{first:b}). Then, to identify $\vect{\lambda}, \vect{\mu}$,   we formulate the following Lagrange dual function:
\begin{equation}
	\begin{aligned}
		g(\vect{\lambda}, \vect{\mu})= \underset{\vect{\pi}}{\text{inf}}\;	\mathcal{L}(\vect{\pi},\vect{\lambda}, \vect{\mu}),
	\end{aligned}
	\label{eq:problem_formulationLagrangian}
\end{equation}
where the solution must satisfy the following  Karush–Kuhn–Tucker (KKT) conditions:
\begin{enumerate}[label=(\roman*)]
	\item Stationarity: ${ \nabla_{\vect{\pi}} }\mathcal{L}(\vect{\pi},\vect{\lambda}, \vect{\mu}) =0;$
	\item Complementary slackness: $ \sum_{v\in \mathcal{V}_r}\lambda_v (x_v^{s\rightarrow r} a^s_{v} - 1)=0$\\ and $\sum_{v\in \mathcal{V}_r}\mu_v(x_v^{s\rightarrow r}p_{vr}y_v^{s\rightarrow r}- P_r)=0;$
	\item Primal feasibility: $\sum_{v\in \mathcal{V}_r}x_v^{s\rightarrow r} a^s_{v}\leq 1$ \\ and $\sum_{v\in \mathcal{V}_r}x_v^{s\rightarrow r}p_{vr}y_v^{s\rightarrow r}\leq P_r;$
	\item Dual feasibility: $\lambda_v \geq0$ and $\mu_v\geq0$.
\end{enumerate}

Our goal is to find the values of 
$\vect{\mu}^*$ and  $\vect{\lambda}^*$ that give the optimal policy $\vect{\pi}^*$ and satisfy the constraints
$ \sum_{v\in \mathcal{V}_r}x_v^{s\rightarrow r} a^s_{v} \leq 1$ and $\sum_{v\in \mathcal{V}_r}x_v^{s\rightarrow r}p_{vr}y_v^{s\rightarrow r}\leq P_r$. However, when $ \sum_{v\in \mathcal{V}_r}\lambda_v (x_v^{s\rightarrow r} a^s_{v} - 1)>0$ and $\sum_{v\in \mathcal{V}_r}\mu_v(x_v^{s\rightarrow r}p_{vr}y_v^{s\rightarrow r}- P_r)>0$,  we can end up paying high penalties $\vect{\mu}$ and  $\vect{\lambda}$. In other words, we  pay high penalties  when resource constraints are violated. To avoid paying such penalties, let us consider a feasible policy $\tilde{\vect{\pi}}$, $\vect{\mu}\succeq 0$, and $\vect{\lambda}\succeq 0$ such that:
\begin{equation}
	\begin{aligned}
		\sum_{r\in \mathcal{R}}\sum_{v\in \mathcal{V}_r}	\tilde{A}^v(\vect{\tilde{\vect{\pi}}}) \geq	\mathcal{L(\tilde{\vect{\pi}},\vect{\lambda}, \vect{\mu}})\geq\underset{\vect{\pi}}{\text{inf}}\;	\mathcal{L}(\vect{\pi},\vect{\lambda}, \vect{\mu}) =g(\vect{\lambda}, \vect{\mu}).
	\end{aligned}
	\label{eq:problem_formulationLagrangian3}
\end{equation}
Therefore, minimizing the overall feasible $\vect{\tilde{\vect{\pi}}}$ gives $\vect{\pi}^* \geq g(\vect{\lambda}, \vect{\mu})$. In other words, the equation  (\ref{eq:problem_formulationLagrangian3}) verifies dual bound theorem. In dual bound theorem, if $\vect{\pi}^* $ is an optimal policy, then
\begin{equation}
	\label{eq:problem_dualbound}
	g(\vect{\lambda}, \vect{\mu}) \leq \sum_{r\in \mathcal{R}}\sum_{v\in \mathcal{V}_r}	\tilde{A}^v(\vect{\vect{\pi}^*}) \leq 	\sum_{r\in \mathcal{R}}\sum_{v\in \mathcal{V}_r}	\tilde{A}^v(\vect{\tilde{\vect{\pi}}}).
\end{equation}
To derive optimal $\vect{\lambda}^*$ and $\vect{\mu}^*$, we can solve the following dual problem:
\begin{subequations}
	\label{eq:problem_formulation1}
	\begin{align}
		&\underset{\vect{\lambda}, \vect{\mu}}{\text{maximize}}\ \  g(\vect{\lambda}, \vect{\mu})
		\tag{\ref{eq:problem_formulation1}}\\
		&\text{subject to:}\nonumber\\
		& \vect{\mu}, \vect{\lambda}\succeq 0,\label{first:aa}\\
		& (\vect{\mu}, \vect{\lambda}) \in \{\vect{\mu}, \vect{\lambda} | g(\vect{\lambda}, \vect{\mu}) > - \infty \}, \label{first:bn}
	\end{align}
\end{subequations}
where the dual problem (\ref{eq:problem_formulation1}) is the lower-bound of the primary problem (\ref{eq:averageAosurrogate}). In (\ref{eq:problem_formulation1}), we exclude $-\infty$ to ensure that the problem (\ref{eq:problem_formulation1})  is lower bound of (\ref{eq:averageAosurrogate}) and feasible.  Therefore, if $\vect{\pi}^*$ is an optimal solution of  $(\ref{eq:averageAosurrogate})$ and $\vect{\lambda}^*$ and $\vect{\mu}^*$  are the solution of (\ref{eq:problem_formulation1}), then
\begin{equation}
	g(\vect{\lambda}^*, \vect{\mu}^*) \leq \sum_{r\in \mathcal{R}}\sum_{v\in \mathcal{V}_r}	\tilde{A}^v(\vect{\vect{\pi}^*}).
\end{equation}
\begin{algorithm}[t]	
	\caption{: AoP-Based Offloading Algorithm.}
	\label{algo:onlienalgorithm}
	\begin{algorithmic}[1]
		\STATE{\textbf{Preconditions:} Each EC knows its CS;}
		\STATE{\textbf{Input:} $\vect{T}$: tasks, $\vect{P}$: computation resources,  $\vect{B}$:  wireless bandwidth, $\omega^s_{v, r}$: fronthaul/Y2 capacity, $\vect{\Gamma}$ link capacity between ECs, and $\omega_{r, RC}$: backhaul capacity;}
		\STATE{\textbf{Output:} Average AoP $\tilde{A}^v(\vect{\vect{\pi}^*})$,  Lagrangian multiplier multipliers $\vect{\mu}^*$ and  $\lambda^*$, Offloading variable $\vect{x}^*$, computation variable $\vect{y}^*$, communication resources allocation  $\vect{a}$, and computation resources allocation $\vect{p}$;}	
		\STATE {Each autonomous vehicle $v \in \mathcal{V}$  chooses  $x_v^{s\rightarrow r}$ and computes $	\alpha_{v}$. When $x_v^{s\rightarrow r}=0$, autonomous vehicle $v \in \mathcal{V}$ computes its task $T_{v}$ locally, calculates $\tau^\textrm{loc}_{v} $, and sets $\tau^\textrm{loc}_{v}=L^\textrm{loc}_{iv}$;}
		\STATE{When $x_v^{s\rightarrow r}=1$, autonomous vehicle $v \in \mathcal{V}$ offloads its task $T_{v}$ to EC $r \in \mathcal{R}$;}
		\STATE {For each task $T_{v}$ reached at any EC $r \in \mathcal{R}$ in CS, EC checks available resources in resource allocation table, calculates $\tau^\textrm{off}_{v}$, and ${re}=x_v^{s\rightarrow r} a^s_{v}$. Then, set $\tau^\textrm{off}_{v} = L^\textrm{off}_{iv}$ ;}
		\STATE{Find the optimal policy $\vect{\pi}*$, $\tilde{A}^v(\vect{\vect{\pi}^*})$, $\vect{\mu}$, $\vect{\lambda}$, $\vect{x}$, and $\vect{y}$ by solving (\ref{eq:problem_formulationLagrangianf}, \ref{eq:problem_formulation1});}
		\STATE{$\vect{x} \leftarrow x_v^{s\rightarrow r}$ and $\vect{a} \leftarrow {re}$, $\vect{p} \leftarrow p_{vr}$ ;}
		\REPEAT
		\STATE{Decrease slightly  $\vect{\mu}$, $\vect{\lambda}$;}
		\STATE{Return to step $7$;}		
		\UNTIL{termination criterion is met (\ref{eq:terminationcondition})}
		\STATE{Then, consider $\vect{x}^*=\vect{x}$, $\vect{y}^*=\vect{y}$, $\vect{\lambda}^*=\vect{\lambda}$ , and $\vect{\mu}^*=\vect{\mu}$ as solution.}
	\end{algorithmic}
\end{algorithm}
To minimize AoP by applying dual decomposition, we propose an age of processing-aware offloading algorithm (Algorithm \ref{algo:onlienalgorithm}). As a precondition of Algorithm \ref{algo:onlienalgorithm}, we assume each EC knows its CS via the Near-RT RIC (Algorithm	\ref{algo:OKM}).

In (\ref{eq:problem_formulationLagrangianf}), we use the penalty Dual Decomposition (DD) method described \cite{shi2020penalty} by integrating Lagrangian multipliers  $\mu_v$ and $\lambda_v$ in the objective function as penalties. According to the convergence analysis of DD provided in \cite{shi2020penalty}, the convergence of Algorithm $2$ is estimated as follows. Let us consider $\{\vect{\pi}^n,\vect{\lambda}^n, \vect{\mu}^n\}$ as the sequences generated in  Algorithm $2$. Here, $\vect{\lambda}^n$ is the sequences generated for the Lagrangian multiplier associated with ($41a$), and $\vect{\mu}^n$ is the sequences generated for the Lagrangian multiplier associated with ($41b$). Furthermore, let   $\epsilon$ be a small positive number, we can formulate the termination criterion for Algorithm $2$ as follows:
	\begin{equation}
		\label{eq:terminationcondition}\lVert e^n\rVert_{\infty}
		\leq \epsilon,   \forall n 
	\end{equation}
	where  $e^n$ is given by:
	\begin{equation}
		\label{eq:termination} e^n=\proj_\mathbf{\pi} \{\vect{\pi}^n - \nabla_{\vect{\pi}}\mathcal{L}(\vect{\pi}^n,\vect{\lambda}^n, \vect{\mu}^n)\} - \vect{\pi}^n.
	\end{equation}
	Let $\vect{\pi}^*$ denote the limit
	point and minimum point  of the sequence  $\{\vect{\pi}^n\}$ when  $n \rightarrow \infty$. Based on dual bound theorem (47), where $\vect{\pi}^* \geq g(\vect{\lambda}, \vect{\mu})$
	and $\vect{\pi}^*$ satisfies KKT stationarity condition for $\nabla_{\vect{\pi}}\mathcal{L}(\vect{\pi}^n,\vect{\lambda}^n, \vect{\mu}^n) =0$. Therefore, for $\nabla_{\vect{\pi}}\mathcal{L}(\vect{\pi}^n,\vect{\lambda}^n, \vect{\mu}^n) =0$, $\lVert e^n\rVert_{\infty}$ goes to zero and this satisfies termination criterion in (\ref{eq:terminationcondition}) and  takes  sublinear convergence $\mathcal{O}\left(\log(1/\epsilon)\right)$.  In  other words, at stationary points $\vect{\pi}^*$ when  $n \rightarrow \infty$,  $\{\vect{\pi}^n\}$ cannot find a better minimum point than $\vect{\pi}^*$. The proof of sublinear convergence of DD is discussed in \cite{necoara2013rate}.

\begin{remark}[The computational complexity of the proposed approach is $O(n^2)$]
	In Algorithm \ref{algo:OKM}, we apply AP. AP has  $O(r^2n)$ computational complexity \cite{refianti2017time}, where $r$ is  the number of data points and $n$ is the number of iterations. In Algorithm \ref{algo:OKM}, the number of data points is $R$, and the number of iterations is $b_m$. Furthermore, in Algorithm \ref{algo:onlienalgorithm}, a task offloaded by vehicle $v$, EC  checks the resource allocation table and allocates resources to the vehicle or forwards the task to another EC.  Since we have $R$ ECs, we assume that checking the resource allocation table (line $6$) takes $n$ iteration. Therefore, checking the resource allocation table has $O(n)$ computational complexity. Furthermore, Algorithm \ref{algo:onlienalgorithm} has a loop (lines $7 -12$) for applying dual decomposition. Considering vehicles $V$, where at each iteration, the Algorithm \ref{algo:onlienalgorithm} has to deal with $\vect{\lambda}$  and $\vect{\mu}$. Consequently,  applying dual decomposition has $O(n^2)$ computational complexity. Therefore, the Algorithm \ref{algo:onlienalgorithm} has $O(n^2+ n)$ computational complexity. From both computation complexities of 
	Algorithm \ref{algo:OKM} and Algorithm \ref{algo:onlienalgorithm}, we conclude that the computational complexity of our proposed approach is $O(n^2)$.
\end{remark} 

In terms of the applicability of our approach, as assumed in \cite{li2021age}, we consider that computation results are smaller than the input data and downlink bandwidth to be larger than uplink. Also, a dedicated channel can be utilized to download computational results. Therefore, the time of transmitting the result to the autonomous vehicle is negligible. Furthermore, the proposed approach has polynomial-time computation complexity $O(n^2)$, where its execution time depends on the number of edge clouds and vehicles. Since many practical problems have polynomial-time solutions \cite{DavidMatuszek}, the proposed approach can be easily implemented in the driving environment. An example of application scenario of our approach is described in the below Section \ref{sec:ApplicationScenario}.
\begin{figure}[t]
	\centering
	\includegraphics[width=1.0\columnwidth]{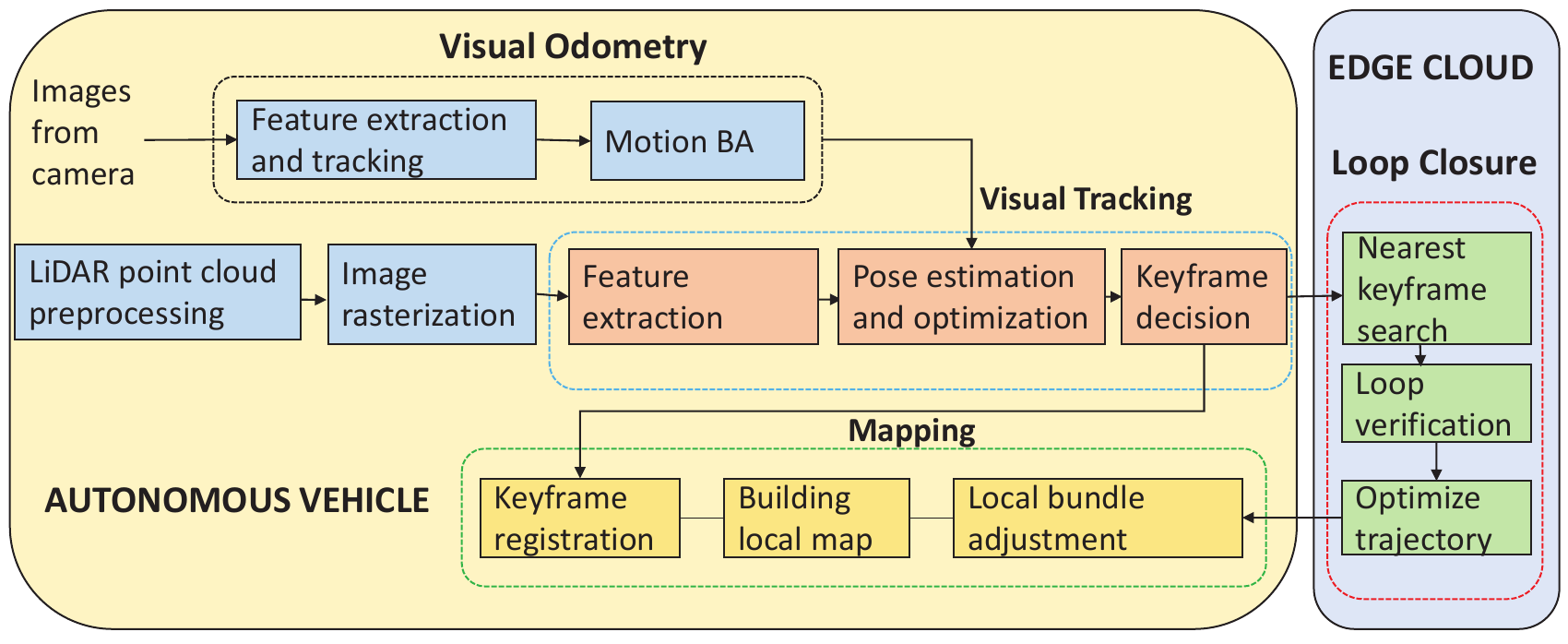}
	\caption{LiDAR-based SLAM system in offloading.}
	\label{LIDAR}
\end{figure}

\subsection{Application Scenario of Age-Based Offloading}
\label{sec:ApplicationScenario}
There are many application scenarios of task offloading for autonomous vehicles.  As an illustrative example, let us consider Simultaneous Localization and Mapping (SLAM) \cite{mur2015orb}. We consider SLAM as a computational problem where an autonomous vehicle builds a map of its current environment. Then, the autonomous vehicle uses the map to navigate the environment. In other words, using sensed data, an autonomous vehicle uses SLAM to generate localized maps. In the implementation, SLAM can use laser sensors.  One of the well-known laser sensors is Light Detection and Ranging (LiDAR), where LiDAR uses pulsed laser waves to map the distance to surrounding objects.

As shown in Fig. \ref{LIDAR}, the LiDAR-based SLAM system  has four modules described in \cite{ali20216} and summarized as follows:
\begin{itemize}
	\item 
	It uses cameras attached to the autonomous vehicle to estimates the egomotion of the autonomous vehicle.  Estimating egomotion works in parallel with the tracking, where image features are extracted and tracked.  Then, motion Bundle-Adjustment (BA)  gives pose in the local frame. Motion BA defines three dimensional coordinates describing the scene geometry relative to motion.
	\item 
	Tracking: Gets rasterized images and extracts features from these	images using Oriented FAST and Rotated BRIEF (ORB). Then,  The tracking thread performs features matching and removes outliers. The matched feature points get projected back to LiDAR coordinates.  Based on the motion transformation, the tracking thread calculates the pose of LiDAR. Then, the tracking thread does a feature consistency checking. The tracking thread can also decide to add a new keyframe. 
	\item 
	Mapping: Once new keyframes are added, the mapping thread registers new keyframes to the keyframe database. Then, simultaneously, the mapping thread builds and saves a local map using pose information. 
	\item 
	Loop Closure: The loop closure thread detects the loop and corrects the accumulated error in the estimated trajectory over time. Loop closure thread searches for the nearest keyframes and performs feature matching to detect loop closure.  Once loop closure is detected,  the loop closure thread builds a pose graph with all keyframes as nodes. Then,  the loop closure thread adds loop closure constraint to the pose graph and optimizes trajectory.
\end{itemize}
Visual Odometry,  Visual Tracking, and Mapping could be executed locally in the autonomous vehicle because they have hard real-time computations. Whereas the loop closure has soft real-time computation \cite{wright2020cloudslam}.  Also, the loop closing thread has a longer execution time and can lead to fast battery power dissipation in vehicle. Therefore, to minimize AoP, the loop closure thread can be offloaded to the ECs for energy-saving and leveraging computational resources of the edge. We assume that the ECs and RC are connected to the power grid and have more computation resources than the autonomous vehicle.

\section{Performance Evaluation}
\label{sec:PerformanceEvaluation}
%++++++++++++++++++++++++++++++++++++++++++++++++++
This section presents the performance evaluation of
the proposed age of processing-based offloading approach for autonomous vehicles.
\begin{figure}[t]
	\centering
	\begin{minipage}{0.45\textwidth}
		\centering
		\includegraphics[width=1.0\columnwidth]{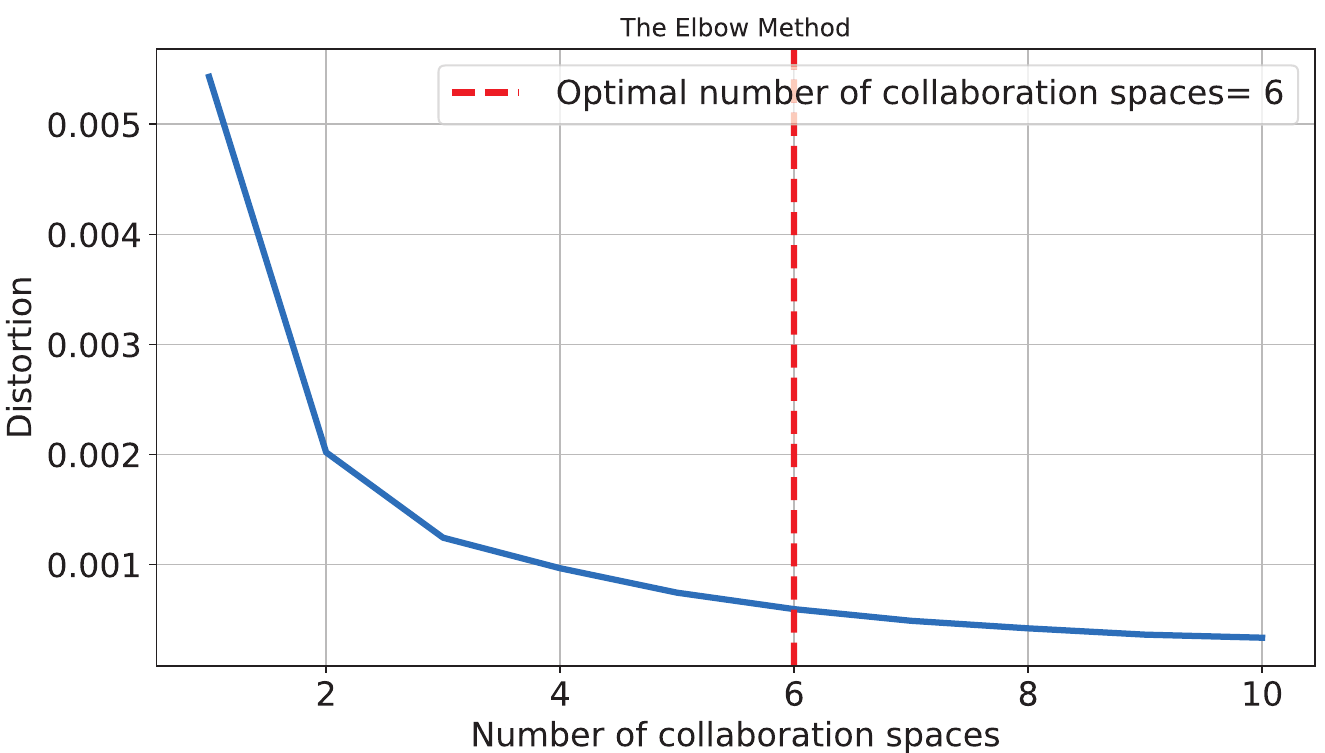}
		\caption{Colloboration space using k-means.}
		\label{fig:kmeans}
	\end{minipage}
	\begin{minipage}{0.45\textwidth}
		\centering
		\includegraphics[width=1.0\columnwidth]{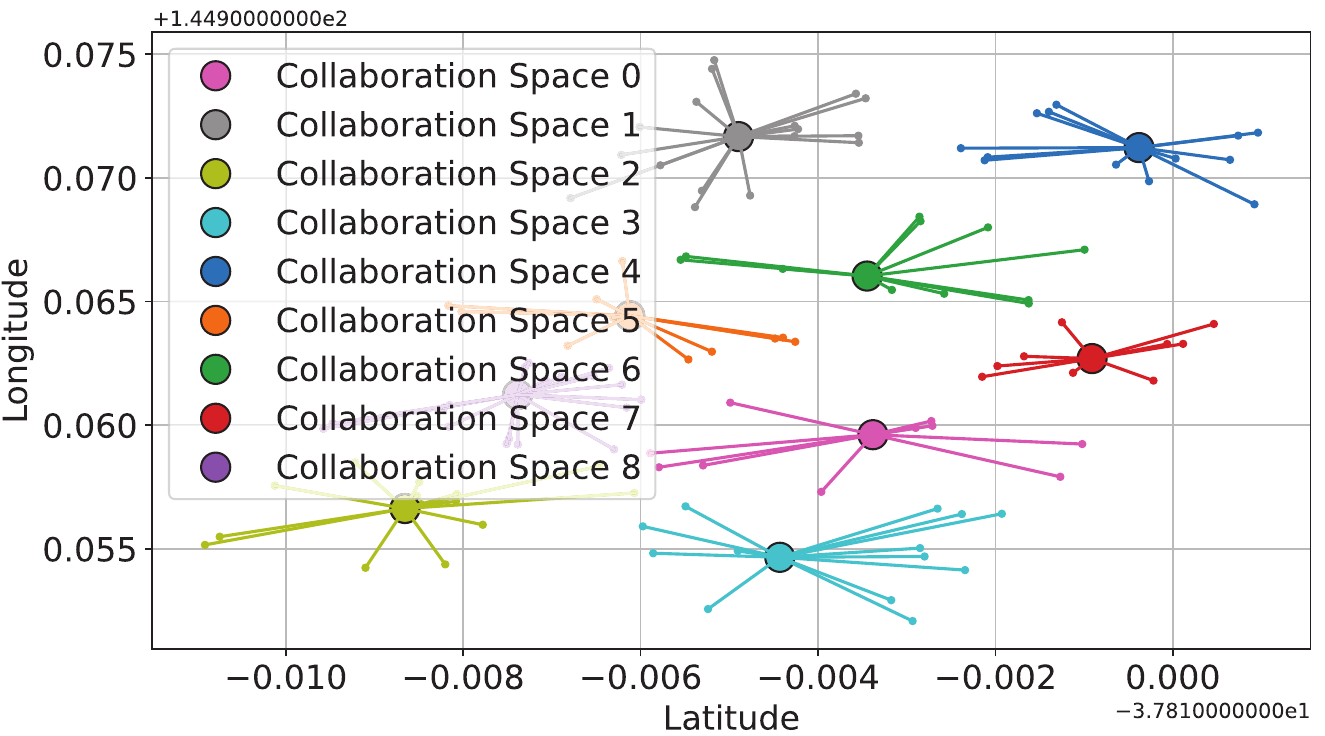}
		\caption{Colloboration space using APACS.}
		\label{fig:APACS}
	\end{minipage}
\end{figure}

\subsection{Simulation Setup}

To form CSs of ECs, we use a distributed computing dataset from the Swinburne University of Technology \cite{lai2018optimal} available at Kaggle \cite{kaggle}. In the spatial coverage of the dataset, we use Melbourne's central business district area. After data preprocessing, in the coverage area, we use $125$ edge servers attached to the base stations.  We consider a single RC that resides outside the spatial coverage of the central business district area. We randomly select $24$ radio base stations  and consider them as RATs. Among these $24$ RATs, we use $6$ as Wi-Fi hotspots, $5$ as RUs, and $13$ as O-RUs.  We generate the autonomous vehicles randomly in a range from $V=5$ to $V=300$. Each autonomous vehicle samples the new status update $i$ and generates one task at each time slot.
For the task $ T_{v}$ of vehicle, the size of the input data $s_{d_v}$ is within a range of $40$ to $200$ $MB$. We randomly generate the task computation deadline of each vehicle $v$  within a range of $\tilde{\tau}_{v}=0.02$ second to $\tilde{\tau}_{v}=1$ second. The computation workload $\tilde{z}_{v}$ of each  vehicle $v$ is in a range  of $\tilde{z}_{v}= 250$ to $\tilde{z}_{v}=9990$ cycles per second. Each autonomous vehicle has a computation resource in the range from $P_v=2.0$ $GHz$ to $P_v=3.0$ $GHz$. Since the dataset does not have computation tasks or resources, we randomly generate the computation tasks and resources. The computation resources of each EC  are in the range from  $P_r=3.0$ $GHz$ to $P_r=3.5$ $GHz$, while at RC, the computation resources are in the range $3.0$ $GHz$ to $4.5$ $GHz$.  

For the communication resources, we set the path loss factor to $4$,  and the transmission power $\varkappa_v=27.0$ dBm. The  cellular channel bandwidth is in the range from  $\omega_{v, s}=25$ MHz to $\omega_{v, s}=32$ MHz \cite{ndikumana2019joint}.  The Wi-Fi bandwidth is  160 MHz (802.11ax) with a maximum theoretical data rate of $\rho^{s}=3.5$ $Gbps$. We consider fronthaul/Y2 bandwidth to be in range  $ \omega_{r, s, t } = 2000$ to $ \omega_{r, s, t }= 2500$ $Mbps$. We set the symmetric bandwidth between each pair of ECs  in the range from $\omega^j_r=3000$ to $\omega^j_r =3500$ $Mbps$. Furthermore, the symmetric bandwidth between each EC and RC is selected in the range from  $\omega_{r, RC} = 3000$ to $\omega_{r, RC}= 4500$ $Mbps$. 

We use Python as a programming language \cite{nagpal2019python} for numerical analysis. For driving route, distance,  and duration, we use the OpenStreetMap routing engine available in \cite{OSRM} and geographic locations of the RATs available in the dataset. Each autonomous vehicle $v$ navigates in the area of $24$ RATs, where speed  $\iota_v$ varies in the range of $4.35$ to $8.63$ meter per second. For the optimization approach, we use CVXPY \cite{diamond2016cvxpy}.
\begin{figure}[t]
	\centering
	\begin{minipage}{0.45\textwidth}
		\centering
		\includegraphics[width=1.0\columnwidth]{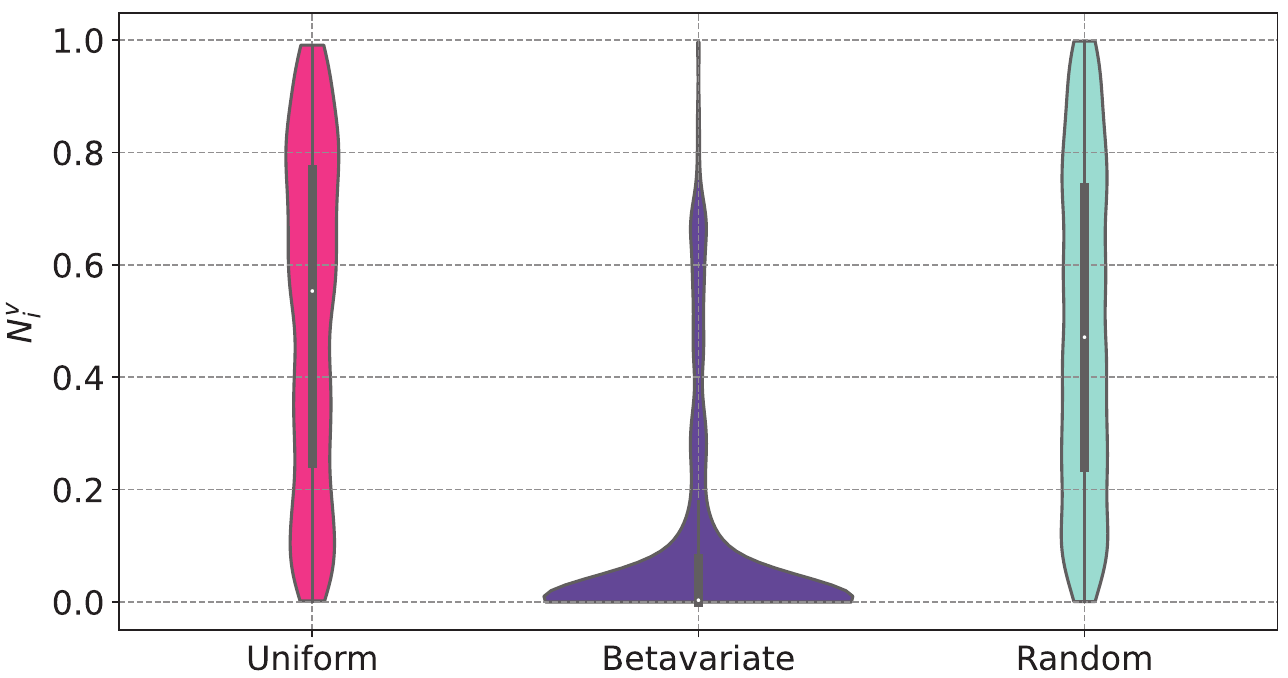}
		\caption{$N^v_i$ for sampling new status update.}
		\label{fig:wait}
	\end{minipage}
	\begin{minipage}{0.45\textwidth}
		\centering
		\includegraphics[width=1.0\columnwidth]{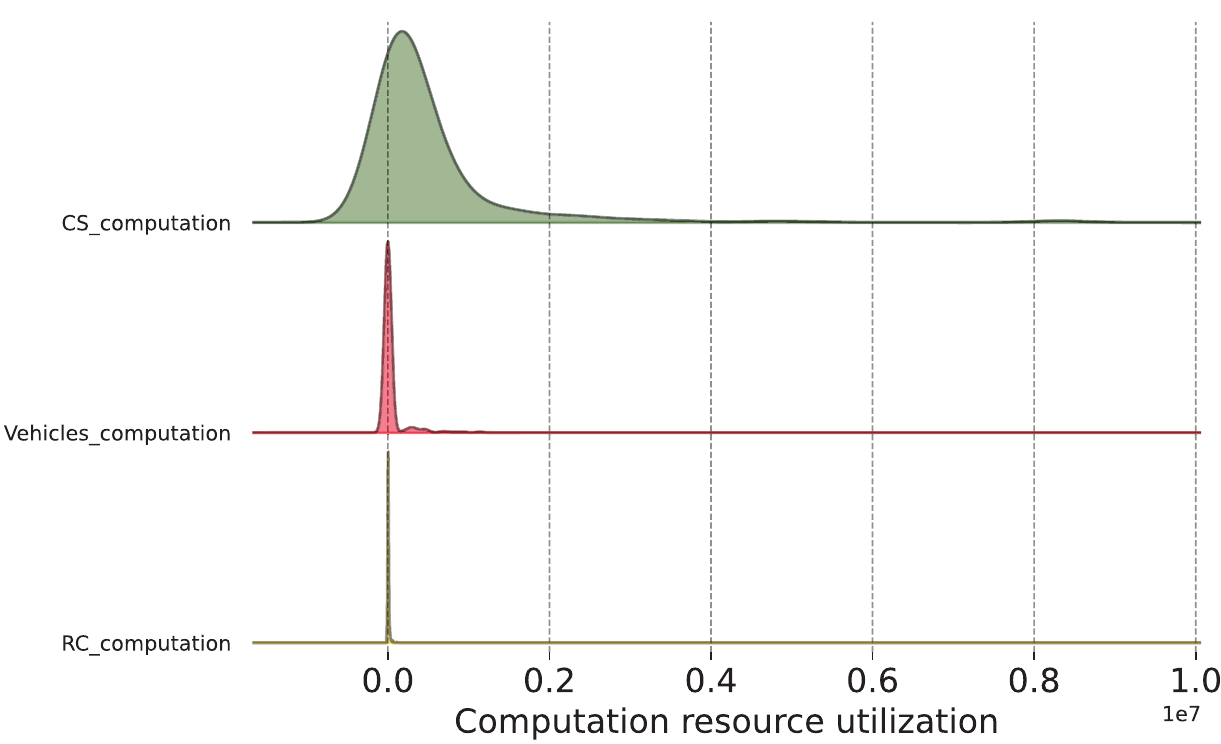}
		\caption{Computation resource utilization.}
		\label{fig:Computation_Resource}
	\end{minipage}
\end{figure}

\subsection{Simulation Results}

In making CSs of ECs, we compare our APACS with k-means. We chose k-means \cite{syakur2018integration} as a baseline over other clustering approaches because APACS and k-means use two different clustering techniques. The k-means uses a fixed number of clusters, while APACS  does not require specifying the number of the clusters, i.e., the number of CSs. APACS takes  measures of similarity between pairs of ECs as input, exchanges messages between ECs,  and gradually emerges ECs into CSs. The simulation results in Fig.
\ref{fig:kmeans} show $6$ CSs as an optimal number of CSs using the Elbow methods and k-means. However, when we use $6$ CSs, some CSs have many ECs (more than $20$
ECs) and this increases the communication delay for exchanging tasks among ECs of the same CS. As shown in \ref{fig:APACS}, to overcome this challenge, we use our APACS. The APACS puts edge ECs in $9$ CSs, where each CS has around $13$ ECs. Since each vehicle $v$ is connected to at least one RAT $s$, RAT $s$ can be connected to EC of any collaboration space among $9$ CSs. We remind that Non-RT RIC at RC runs APACS to make CSs. Then, Non-RT RIC informs each EC about its CS via Near-RT RIC. Since ECs' network topology does not change frequently, we also assume that CSs do not change frequently. This motivates us to use two algorithms:  Algorithm 1 (APACS), which runs at Non-RT RIC for the formation of CSs; and Algorithm 2 (AoP-Based Offloading Algorithm), which runs in both EC and vehicle for AoP-based offloading. Furthermore, ECs exchange resource utilization information and tasks to help each other in computing tasks as close as possible to autonomous vehicles for minimizing AoP.  In other words, Algorithm 2 deals with the topology that frequently changes because of the vehicle's connection in motion and high mobility.
\begin{figure}[t]
	\centering
	\begin{minipage}{0.45\textwidth}
		\centering
		\includegraphics[width=1.0\columnwidth]{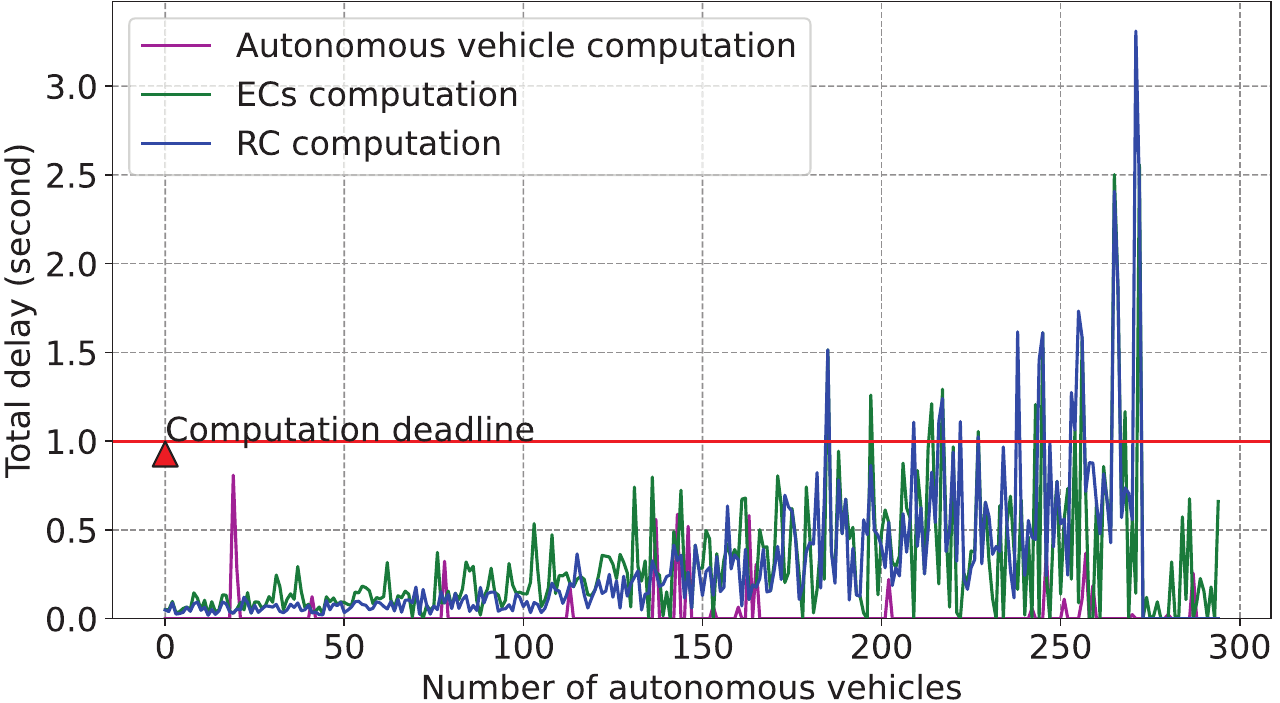}
		\caption{Total delay before optimization.}
		\label{fig:total_delay1}
	\end{minipage}
	\begin{minipage}{0.45\textwidth}
		\centering
		\includegraphics[width=1.0\columnwidth]{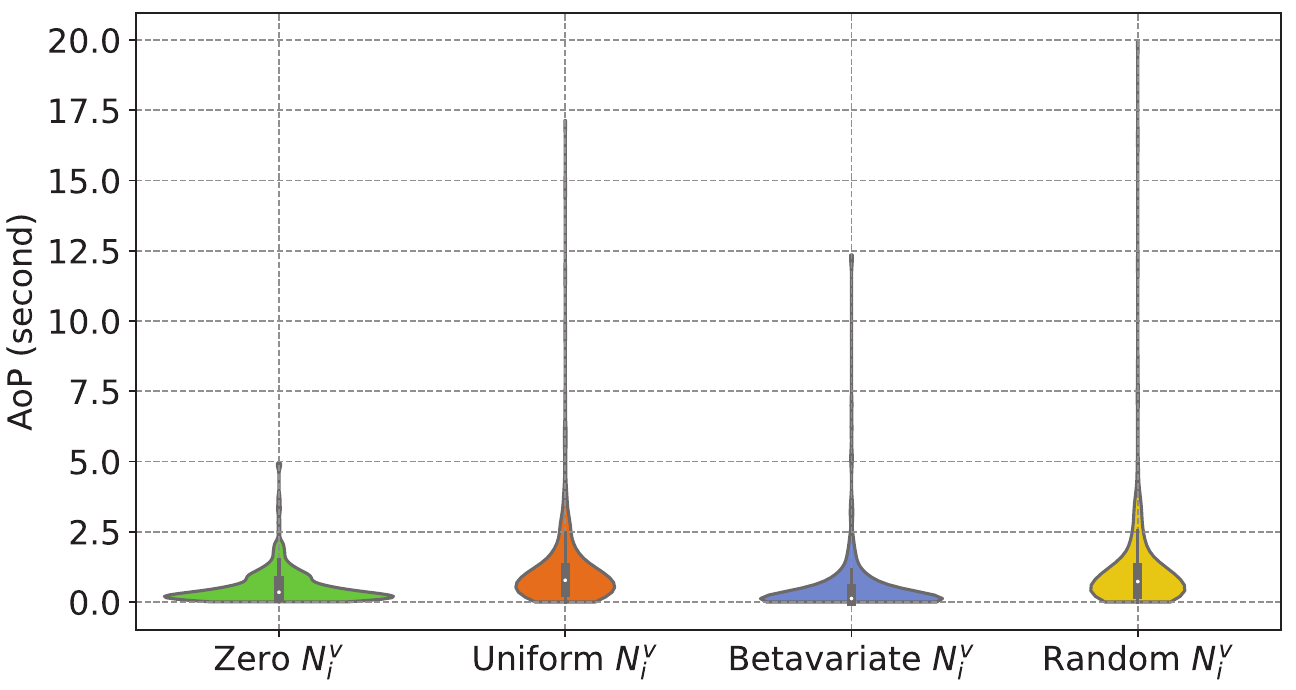}
		\caption{AoP before optimization.}
		\label{fig:AoP}
	\end{minipage}
\end{figure}
\begin{figure}[t]
	\centering
	\begin{minipage}{0.45\textwidth}
		\centering
		\includegraphics[width=1.0\columnwidth]{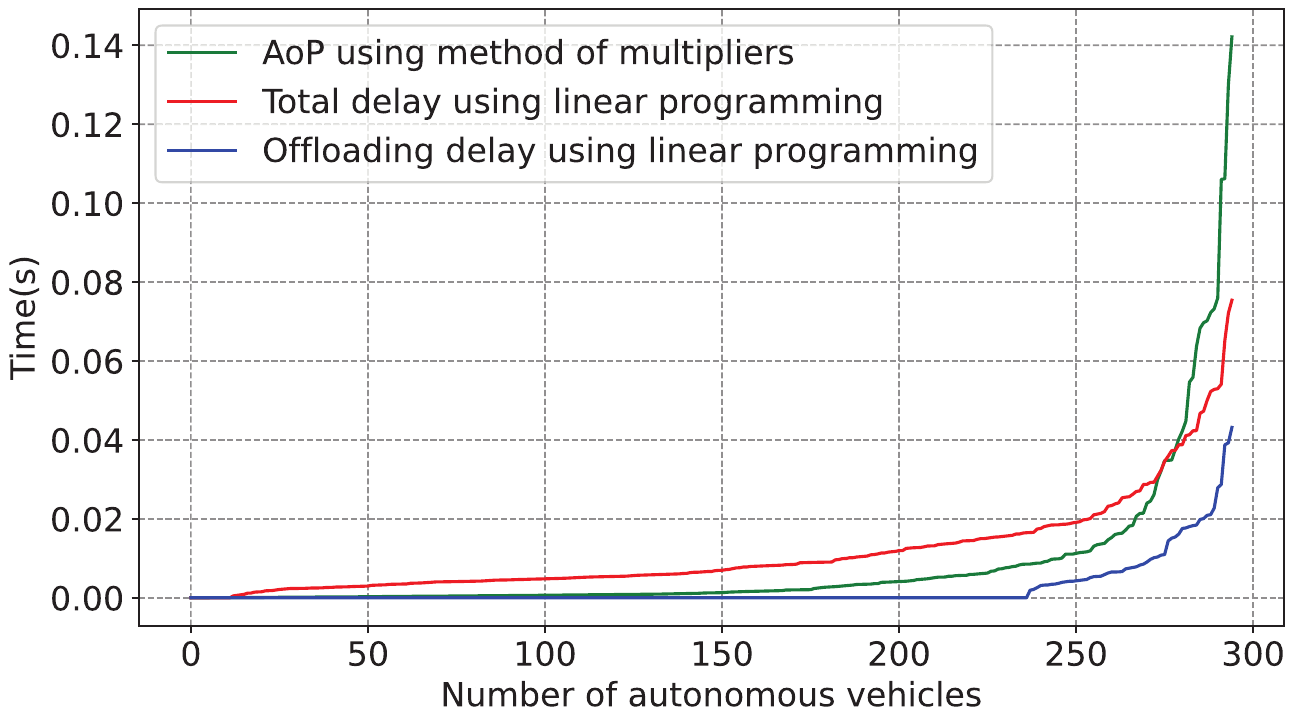}
		\caption{Total delay versus AoP.}
		\label{fig:AoP_TotalDelay}
	\end{minipage}
	\begin{minipage}{0.45\textwidth}
		\centering
		\includegraphics[width=1.0\columnwidth]{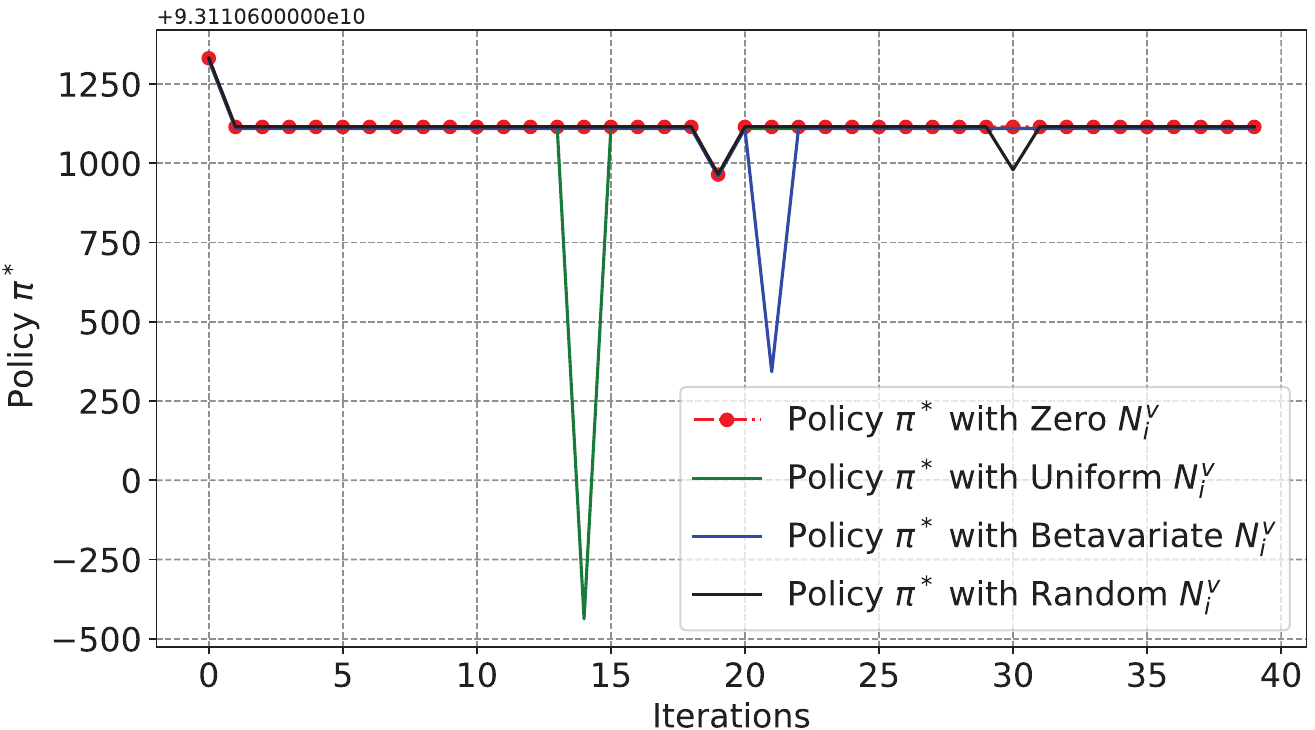}
		\caption{Computation policy $\pi^*$.}
		\label{fig:Policy}
	\end{minipage}
\end{figure}

We consider that the  vehicle samples new status update $i+1$ after time $N^v_i \geq 0$. In Fig. \ref{fig:wait}, we use zero wait for $N^v_i=0$, where the vehicle continuously keeps sampling new status updates. We also consider random, uniform, and betavariate $N^v_i$ for sampling new status update $i+1$, where $N^v_i$ is in the range between $0$ and $1$ second. 
\begin{figure}[t]
	\centering
	\begin{minipage}{0.45\textwidth}
		\centering
		\includegraphics[width=1.0\columnwidth]{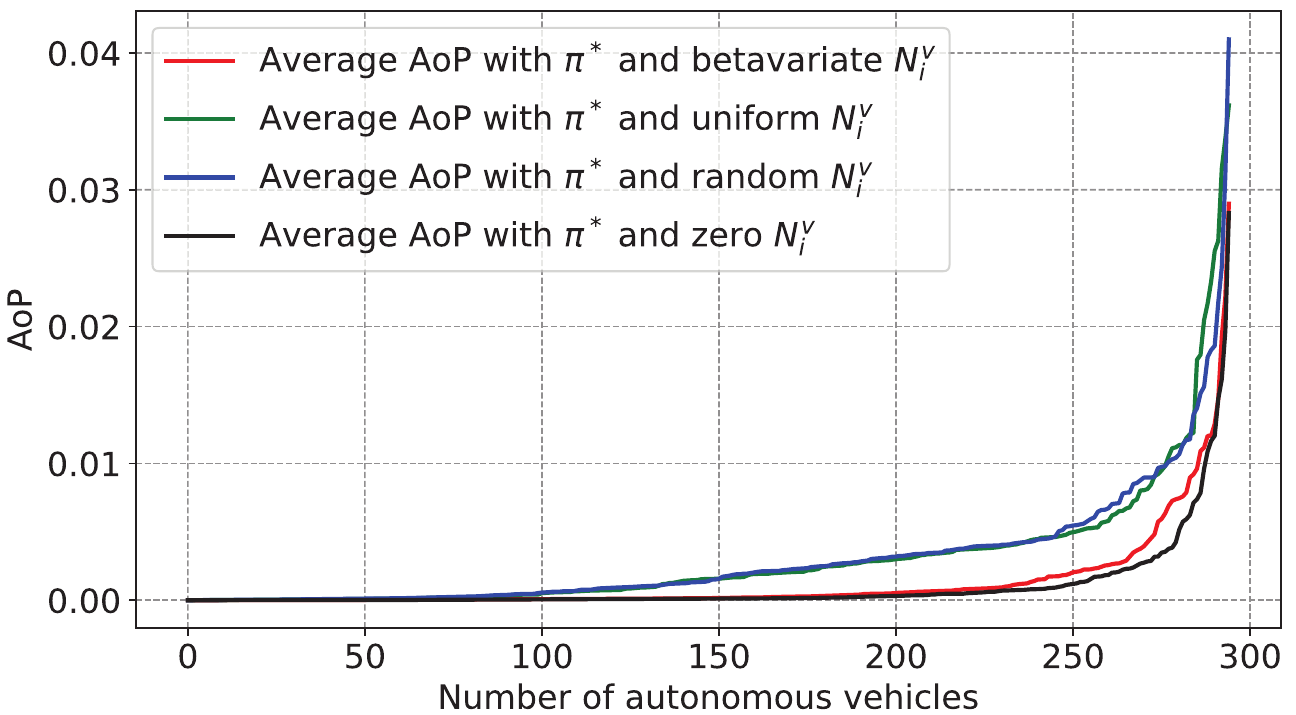}
		\caption{AoP with $\pi^*$ and different $N^v_i$ .}
		\label{fig:AoP_Optimal}
	\end{minipage}
	\begin{minipage}{0.45\textwidth}
		\centering
		\includegraphics[width=1.0\columnwidth]{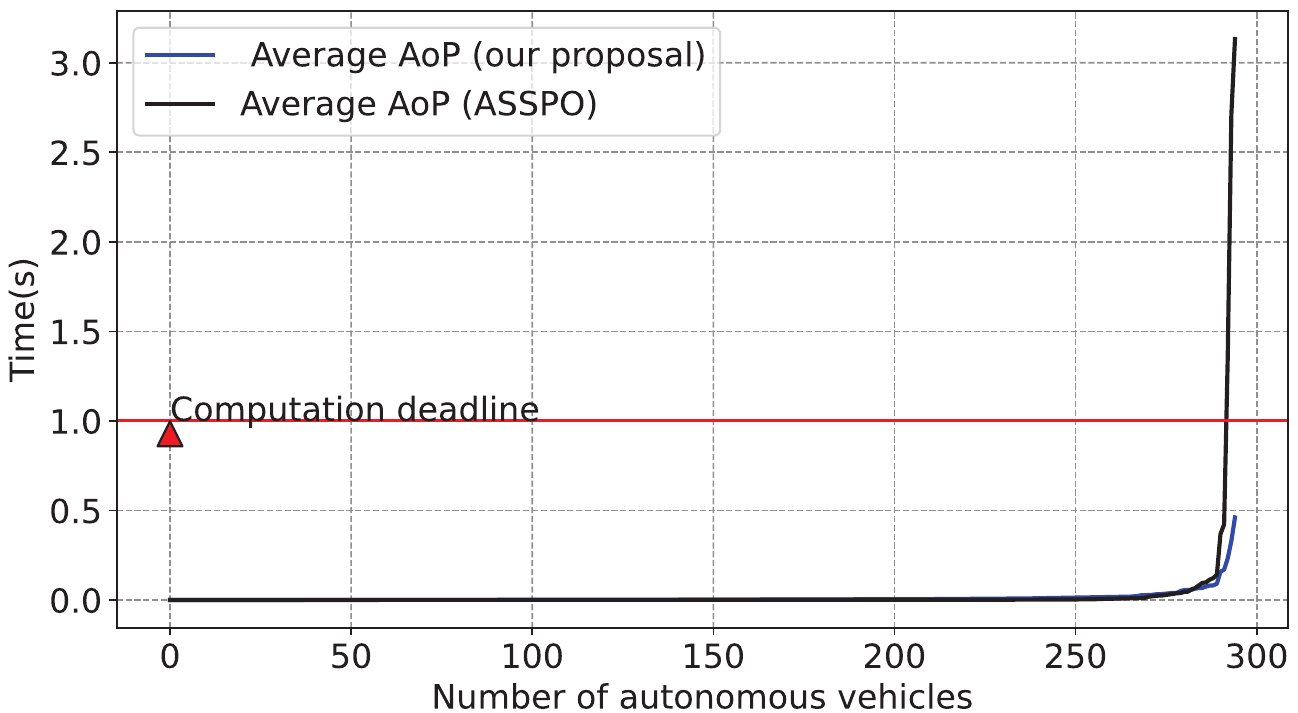}
		\caption{Our approach vs ASSPO.}
		\label{fig:compare}
	\end{minipage}
\end{figure}
Furthermore, Fig. \ref{fig:Computation_Resource} shows computation resources utilization. Most of the computations are performed in CS, i.e.,  at ECs because the vehicles have limited resources. There is a collaboration of ECs of the same CS to avoid sending more tasks to a remote RC at a far transmission distance. Fig. \ref{fig:total_delay1} shows the comparison of delays, where local computation in the autonomous vehicle meets the computation deadline because local computation does not involve offloading time. However, offloading tasks to the edge and regional clouds may violate the computation deadline due to transmission and propagation delays. In considering $N^v_i$, we compute AoP. The results in Fig. \ref{fig:AoP} show that the $N^v_i=0$ has the best performance compared to other values of $N^v_i$.  When the vehicle keeps $N^v_i$ very close to $zero$ for generating new update $i+1$, each generated input data is small, and offloading input data takes less time. In random and uniform $N^v_i$, when $N^v_i$ is very close to $1$ second, input data becomes large, and offloading input data takes more time. In such a case, the AoP becomes large, and the computation cannot meet the deadline because 1 second is quite large for autonomous driving.

After applying our optimization approach, in Fig \ref{fig:AoP_TotalDelay}, we compare the total delay (offloading and computation delay) and AoP  with zero $N^v_i$. Also, in this figure, we show the offloading delay. The gap between offloading delay and total delay/AoP is the computation delay. The optimization approach helps to ensure offloading and computation meet the deadline.   The simulation results show that using AoP achieves better performance than minimizing the total delay. In the total delay, we solve (\ref{eq:problem_formulation}) by considering (\ref{eq:total_compution_cost}) as an objective function. 
Furthermore, we compute the optimal policy $\vect{\pi}^*$ that minimizes the AoP and show the result in Fig. \ref{fig:Policy}. In other words, this figure shows the convergence of (\ref{eq:averageAosurrogate}). Considering policy $\vect{\pi}^*$ that prevents the violation of computation and communication constraints, Fig. \ref{fig:AoP_Optimal} shows average AoP with $\pi^*$ and different $N^v_i$. 

In our approach, the vehicle senses the environment. Then, it uses the status update to make safe and reliable autonomous driving decisions without relying on external operators to validate the status update. In other words, a vehicle plays the roles of operator and sensing node simultaneously. This consideration has not been tackled so far in the existing AoP or AoI-based offloading approaches. However, we compared our approach with the AoP model presented in \cite{li2021age} (denoted ASSPO in Fig \ref{fig:compare}). In the ASSPO model, we consider an operator at the RC that controls the sensed data of edge device (i.e., a vehicle in our approach). The operator sends an acknowledgment for each received status update. Then,   ASSPO calculates AoP based on the reception of acknowledgment. The simulation results in Fig. \ref{fig:compare} show that our approach achieves a  lower AoP than that of ASSPO because the vehicle does not have to wait for the acknowledgment before sampling a new status update. In other words, ASSPO always has to wait for the acknowledgment before sampling a new status update.

\section{Conclusion}
\label{sec:Conclusion}
To meet the computation deadline of autonomous vehicle, we propose an offloading approach that supports OBU and enables the autonomous vehicle to offload computation tasks to the edge clouds. In the proposed offloading approach, the CS of edge clouds guarantees that vehicles' tasks are computed as closely as possible to the vehicles. Since the network status changes over time,  to achieve less variation in delay for offloading tasks, our new communication planning approach enables the vehicle to preselect appropriate RATs available in its route to use for offloading tasks. The simulation results clearly show that our proposed approach satisfies computation deadlines by minimizing AoP. This work focuses on AoP and delay as metrics. In future work, we plan to enhance our age-based offloading for autonomous vehicles in a dynamic network environment and consider more metrics such as reliability, throughput, packet loss, and retransmission.
\section*{Acknowledgment}
\noindent The authors thank Mitacs, Ciena, and ENCQOR for funding this research under the IT13947 grant.
\bibliographystyle{IEEEtran}
	
\begin{IEEEbiography}[{\includegraphics[width=1in,height=1.25in,clip,keepaspectratio]{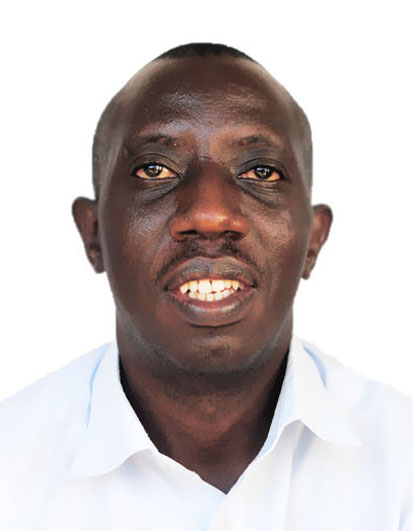}}]{Anselme Ndikumana} received  B.S. degree in Computer Science from the National University of Rwanda in 2007 and Ph.D. degree in Computer Engineering from Kyung Hee University, South Korea in August 2019. Since 2020, he has been	with the Synchromedia Lab, École de Technologie Supérieure, Université du Québec, Montréal, QC, Canada where he is currently a postdoctoral fellow. His professional experience includes Lecturer at the University of Lay Adventists of Kigali from 2019 to 2020, Chief Information System, a System Analyst, and a Database Administrator at Rwanda Utilities Regulatory Authority from 2008 to 2014. His research interest includes AI for wireless communication, multi-access edge computing, 5G networks, information-centric networking, and in-network caching.
\end{IEEEbiography}
\begin{IEEEbiography}[{\includegraphics[width=1in,height=1.25in,clip,keepaspectratio]{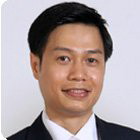}}]{Kim Khoa Nguyen} is Associate Professor in the Department of Electrical Engineering and the founder of the Laboratory on IoT and Cloud Computing at the University of Quebec’s  Ecole de technologie superieure. In the past, he served as CTO of Inocybe Technologies (now is Kontron Canada), a world’s leading company in software-defined networking (SDN) solutions. He was the architect of the Canarie’s GreenStar Network and led R\&D in large-scale projects with Ericsson, Ciena, Telus, InterDigital, and Ultra Electronics. He is the recipient of Microsoft Azure Global IoT Contest Award 2017, and Ciena’s Aspirational Prize 2018. He is the author of more than 100 publications, and holds several industrial patents. His expertise includes network optimization, cloud computing IoT, 5G, big data, machine learning, smart city, and high speed networks.
\end{IEEEbiography}
\begin{IEEEbiography}[{\includegraphics[width=1in,height=1.25in,clip,keepaspectratio]{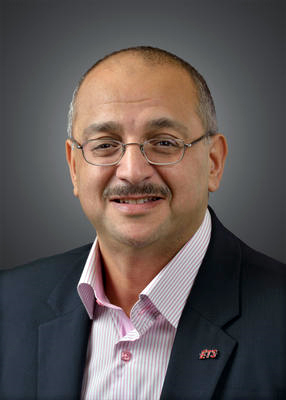}}]{Dr. Mohamed Cheriet} received his Bachelor, M.Sc. and Ph.D. degrees in Computer Science from USTHB (Algiers) and the University of Pierre \& Marie Curie (Paris VI) in 1984, 1985 and 1988 respectively. He was then a Postdoctoral Fellow at CNRS, Pont et Chaussées, Paris V, in 1988, and at CENPARMI, Concordia U., Montreal, in 1990. Since 1992, he has been a professor in the Systems Engineering department at the University of Quebec - École de Technologie Supérieure (ÉTS), Montreal, and was appointed full Professor there in 1998. Prof. Cheriet was the director of LIVIA Laboratory for Imagery, Vision, and Artificial Intelligence (2000-2006), and is the founder and director of Synchromedia Laboratory for multimedia communication in telepresence applications, since 1998.  Dr. Cheriet research has extensive experience in Sustainable and Intelligent Next Generation Systems. Dr. Cheriet is an expert in Computational Intelligence, Pattern Recognition, Machine Learning, Artificial Intelligence and Perception and their applications, more extensively in Networking and Image Processing. In addition, Dr. Cheriet has published more than 500 technical papers in the field and serves on the editorial boards of several renowned journals and international conferences. He held a Tier 1 Canada Research Chair on Sustainable and Smart Eco-Cloud (2013-2000), and lead the establishment of the first smart university campus in Canada, created as a hub for innovation and productivity at Montreal. Dr. Cheriet is the General Director of the FRQNT Strategic Cluster on the Operationalization of Sustainability Development, CIRODD (2019-2026). He is the Administrative Director of the \$12M CFI’2022 CEOS*Net Manufacturing Cloud Network. He is a 2016 Fellow of the International Association of Pattern Recognition (IAPR), a 2017 Fellow of the Canadian Academy of Engineering (CAE), a 2018 Fellow of the Engineering Institute of Canada (EIC), and a 2019 Fellow of Engineers Canada (EC). Dr. Cheriet is the recipient of the 2016 IEEE J.M. Ham Outstanding Engineering Educator Award, the 2013 ÉTS Research Excellence prize, for his outstanding contribution in green ICT, cloud computing, and big data analytics research areas, and the 2012 Queen Elizabeth II Diamond Jubilee Medal. He is a senior member of the IEEE, the founder and former Chair of the IEEE Montreal Chapter of Computational Intelligent Systems (CIS), a Steering Committee Member of the IEEE Sustainable ICT Initiative, and the Chair of ICT Emissions Working Group. He contributed 6 patents (3 granted), and the first standard ever, IEEE 1922.2, on real-time calculation of ICT emissions, in April 2020, with his IEEE Emissions Working Group.
	
\end{IEEEbiography}
\end{document}